\documentclass{aa}  

\usepackage{graphicx}
\usepackage{txfonts}
\usepackage{lipsum}
\usepackage{subcaption}         
\usepackage{lscape}             
\usepackage{placeins}           
\usepackage{hyperref}
\usepackage{float} 
\usepackage[normalem]{ulem}
\usepackage[dvipsnames]{xcolor}

\begin{document}

   \title{Pulsar Electrodynamics Inferred from Frequency-Dependent Circular Polarization Diversity}

   \author{Shunshun Cao\inst{1}\fnmsep\inst{2}, Yanjun Guo\inst{3}\fnmsep\inst{4}\fnmsep\thanks{guoyj@bao.ac.cn}, Jinchen Jiang\inst{5}\fnmsep\inst{4}\fnmsep\inst{2}, Kejia Lee\inst{2}\fnmsep\inst{6}\fnmsep\inst{4}, Weiyang Wang\inst{7}, \and Renxin Xu\inst{1}\fnmsep\inst{2}\fnmsep\inst{6}\fnmsep\thanks{r.x.xu@pku.edu.cn}}

   \institute{State Key Laboratory of Nuclear Physics and Technology, Peking University, Beijing 100871, China \and Department of Astronomy, School of Physics, Peking University, Beijing 100871, China
            \and State Key Laboratory of Radio Astronomy and Technology, National Astronomical Observatories, Chinese Academy of Sciences, Beijing 100101, China \and National Astronomical Observatories, Chinese Academy of Sciences, Beijing 100012, China \and Max-Planck-Institut f\"ur Radioastronomie, Auf dem H\"ugel 69, D-53121 Bonn, Germany \and Kavli Institute for Astronomy and Astrophysics, Peking University, Beijing 100871, China \and School of Astronomy and Space Science, University of Chinese Academy of Sciences, Beijing 100049, China}

   \date{Received;}

  \abstract 
   {The nature of coherent radio emission remains challenging even after more than half a century of pulsar discovery. However, the general consensus asserts that single-pulse observations are essential for probing the magnetospheric dynamics, especially with the largest single-dish telescope presently available: Five-hundred-meter Aperture Spherical radio Telescope (FAST).}
   {This paper is aimed at explaining the observed diversity of single pulse circular polarization and constraining the multiplicity and Lorentz factor of pulsar magnetospheric plasma, with the mode coupling model in the limiting polarization region.} 
   {Assuming that circular polarization only comes from wave mode coupling, we applied a Bayesian analysis to the FAST observed single pulse circular polarization spectra, including a numerical solution for the wave-mode coupling equations. We also analyzed the posterior probability distribution functions of the parameters.}
   {Although the model failed to quantitatively fit most circular polarization spectra, the circular polarization of different frequency evolution was successfully reproduced. For the three chosen pulsars, the Bayesian analysis constrained the multiplicity as approximately $10^{0}\sim10^{2}$, along with a Lorentz factor of approximately $10^{0.5}\sim10^{2}$.}
   {Pulsar circular polarization could be induced by wave mode coupling. The plasma flow responsible for coherent radio emission carries only a very small fraction of the pulsar spin-down energy loss.}


   \maketitle
   \nolinenumbers

\section{Introduction}\label{sec:intro}
Even after nearly six decades since the discovery of pulsars, the physics of cosmic coherent radio emission is not fully understood. This is likely due to the nature of pulsars, particularly with regard to pulsar surfaces~\citep[e.g.,][]{xuzihao2025}. New hot topics, such as fast radio bursts~\citep{sironi2021, wwy2022} and long-period radio transients~\citep{qz2025,zm2025}, could make this puzzling problem even more complex. Nevertheless, advanced facilities such as China's FAST help to resolve these big questions by providing high-quality pulsar data for probing pulsar magnetospheres. In these environments, the physical processes have not been fully clarified and the related parameters have not been determined with a high level of confidence. The circular polarization behaviors of single pulses detected by FAST are the focus of this paper to aid in our understanding of pulsar electrodynamics.

Linearly polarized radiation predominates in the pulsar magnetosphere, as coherent radio emission typically arises from one of two orthogonal linear polarization modes. Consequently, the circular polarization component is essential for understanding the underlying radiation mechanisms.

The observed circular components is, in fact closely, associated with pulsar magnetospheric properties and its generation mechanism is based on the modeling of emission mechanisms or wave propagation processes~\citep[e.g.,][]{ab1986,kazbegi1991,lp1998,melrose2004,benacek2025}. Observational studies have already reported a great diversity among the circular polarization phenomenology of pulsar integrated profiles~\citep{han1998,you2006}, pulsar single pulses~\citep{karastergiou2003}, and fast radio bursts~\citep{jjc2024}. However, the comparisons between observations and theories are usually qualitative (or just on an order of magnitude), which stands in the way of an effective elimination of unhelpful models. A quantitative comparison between pulsar emission models and observational data would not only help us better understand the advantages and disadvantages of certain models, but it would also provide plausible ways of diagnosing pulsar magnetospheric plasmas, as some pioneer research has shown~\citep{petrova2003,galishnikova2020}. These studies have been carried out on the basis of integrated profiles and we hope to achieve good-quality diagnostics with single pulses.

In this paper, we use the mode coupling model in the limiting polarization region of pulsar magnetosphere to fit the longitude-resolved circular polarization spectra of single pulses observed by FAST. The role of the limiting polarization effect has been highlighted in studies of radio pulsar polarization patterns~\citep{cr1979, lp1998, wangchen2010, ab2010, bp2012}. Certainly, the wave modes in strongly magnetized plasma are highly linear polarized~\citep[e.g.,][]{ab1986}. However, in the limiting polarization region of a magnetosphere, the inhomogeneity of the ambient plasma is strong enough to make the wave modes couple with each other during propagation. When the magnetic field orientation changes along the ray trajectory, the original linear polarization of one single wave mode fails to catch up with the magnetic field, which results in the emergence of another wave mode and finally induces a circular polarization component in the outgoing wave.

The paper is structured as follows. In Section~\ref{sec:model}, we describe the wave mode coupling model in the limiting-polarization region of the pulsar magnetosphere. We clarify the physical quantities and the assumptions of the model. Section~\ref{sec:data} contains the information of the FAST data we use and how we process the data. In Section~\ref{sec:computational}, we introduce our settings for the computational methods, including Bayesian analysis and numerical integration. Section~\ref{sec:results} gives the results of Bayesian analysis. Section~\ref{sec:discussions} presents a discussion of the results, including the implications for magnetospheric dynamics and particle accelerations. Section~\ref{sec:conclusions} concludes our findings.

\section{The wave mode coupling model}\label{sec:model}
    The polarization transfer equations of waves propagating in astrophysical plasmas have been studied extensively ~\citep[e.g.,][]{sazonov1968,jo1977}. The wave-mode coupling in an inhomogeneous plasma, as an example, was first studied for the Earth ionosphere~\citep[e.g.,][]{budden1952} and later applied to investigations of astrophysical phenomena~\citep[e.g.,][]{zhbook}. \cite{budden1952} used the parameter $\psi$ to describe how large the coupling is
    \begin{equation}
        \psi=\dfrac{i}{R_{\mathrm{O}}^{2}-1}\dfrac{dR_{\mathrm{O}}}{dh}
        \label{eq:limiting_pol_general},
    \end{equation}
    \noindent where $R_{\mathrm{O}}=E_{\mathrm{O},y}/E_{\mathrm{O},x}$ describe the polarization of the ordinary (O) mode and $h$ is a coordinate along the ray trajectory. A critical value of $\psi$ is $k|n_{\mathrm{O}}-n_{\mathrm{E}}|/2$, where $k$ is the wave number, while $n_{\mathrm{O,E}}$ represents the refraction indices of the O mode and extraordinary (E) mode. When $\psi$ increases, coupling becomes significant. In the case of pulsar magnetosphere, $\psi$ is on the order of $1/r$, where $r$ is the distance from neutron star centroid. So the criterion for mode coupling could be converted to
    \begin{equation}
        \dfrac{\omega}{c}|n_{\mathrm{O}}-n_{\mathrm{E}}|r\sim 1
        \label{eq:limiting_pol_psr}.
    \end{equation}
    Equation ~\ref{eq:limiting_pol_psr} defines a limiting polarization radius, $r_{\mathrm{pl}}$, which has been discussed in detail by \cite{barnard1986}.
    
    For the mode coupling equations, we used the formulae established by \cite{lp1998} (i.e., Budden-Zheleznyakov approach in \citealp{bp2012}), expressed as
    \begin{equation}
			\left\{
			\begin{aligned}
				\dfrac{da_{x}}{dz}&=-iR[(b_{x}+q_{y})^{2}a_{x}+(b_{x}+q_{y})(b_{y}-q_{x})a_{y}],\\
				\dfrac{da_{y}}{dz}&=-iR[(b_{x}+q_{y})(b_{y}-q_{x})a_{x}+(b_{y}-q_{x})^{2}a_{y}],
			\end{aligned}
			\right.
            \label{eq:mode_coupling}
	\end{equation}
   where the parameter $R$ contains the plasma parameters via
    \begin{equation}
        R = \dfrac{\omega_{p}^{2}}{2\omega c\gamma^{3}(1-\beta_{z})^{2}}=\dfrac{4\pi\kappa n_{\mathrm{GJ}}(z)e^{2}/m_{e}}{\omega c \gamma^{3}(1-(\mathbf{\Omega}\times \mathbf{r})_{z}/c-v_{b}b_{z}/c)^{2}}
        \label{eq:R_para}.
    \end{equation}
    In Eq.~\ref{eq:mode_coupling} and Eq.~\ref{eq:R_para}, the $z$ axis is chosen to be along the ray trajectory. The meanings of physical quantities are presented in Table~\ref{table:quantities}.

    \begin{table}[h!]
    \caption{Physical quantities in Eq.~\ref{eq:mode_coupling} - Eq.~\ref{eq:q_vec}}                
    \label{table:quantities}   
    \centering                        
    \begin{tabular}{c l}     
    \hline\hline               
    Quantity & Meaning\\         
    \hline                      
    $\mathbf{a}$ & complex electric vector \\    
    $\mathbf{b}$ & direction vector of magnetic field  \\
    $\mathbf{q}$ &  $\mathbf{b}\times(\mathbf{\Omega}\times \mathbf{r})/c$ \\
    $\mathbf{\mathbf{\Omega}}$ & angular velocity of pulsar rotation\\
    $\omega$ & angular frequency of the propagating wave  \\
    $\gamma$ & Lorentz factor of $e^{\pm}$  \\
    $v_{b}$ & $e^{\pm}$ velocity component along magnetic field lines \\
    $n_{\mathrm{GJ}}$ & $|\rho_{\mathrm{GJ}}/e|$, Goldreich-Julian plasma number density~\\ & \citep{gj1969} \\
    $\kappa$ & multiplicity \\
    $m_{e}$ & electron mass\\
    $c$ & vacuum speed of light \\
    $\mathbf{\mu}$ & magnetic dipole moment of the pulsar \\
    $\alpha$ & inclination angle \\
    $\zeta$ & angle between $\mathbf{\Omega}$ and $\mathbf{k}$\\
    $\delta$ & angle between $\mathbf{\mu}$ and $\mathbf{k}$ \\
    $z_{0}$ & starting point of integration \\
    $z_{1}$ & ending point of integration \\
    \hline                                  
    \end{tabular}
    \end{table}
    
    The derivation of Eq.~\ref{eq:mode_coupling} could be found in \cite{lp1998} and \cite{pl2000}. This form of coupling equations are valid under the following assumptions:
    \begin{enumerate}
        \item Highly relativistic plasma: $\gamma^{2}\gg 1$;
        \item High frequency waves: $Rc/\omega\ll 1$ (similar discussions are made in \citealp{bgi1988});
        \item The plasma is cold, with single Lorentz factor;
        \item The plasma is strictly frozen on the magnetic field and the velocity can be written as $\mathbf{v}=v_{b}\mathbf{b}+\Omega\times \mathbf{r}$;
        \item Refraction is neglected in the limiting polarization region ($n_{\mathrm{O,E}}\sim 1$).
    \end{enumerate}

    In \cite{lp1998}, $\mathbf{b}$ and $\mathbf{q}$ could be derived in a simple form under two extra assumptions for the magnetic field and geometrical parameters:

    \begin{enumerate}
        \item $\delta\ll\zeta$;
        \item The region of interest (limiting polarization region) is far enough from the neutron star surface, and rotation dominates the change of magnetic field along the ray trajectory.
    \end{enumerate}

   Following \cite{lp1998}, we set the $x$ axis along the initial $\mathbf{b}-(\mathbf{b}\cdot\mathbf{k}/k)\mathbf{k}$, in other words, we set the initial $b_{y}=0$. The initial $\mathbf{b}$ is $(\delta/2,0,1-\delta^{2}/8)$, the orientation of rotation axis is $\hat{\Omega}=(\sin\zeta \sin\theta, \sin\zeta \cos\theta, \cos\zeta)$, and the orientation of magnetic dipole is $\hat{\mu}=(-\sin\delta,0,\cos\delta)$. The angle $\theta$ satisfies $\hat{\Omega}\cdot\hat{\mu}=\cos{\alpha}$. Then the rotation matrices are introduced as

   \begin{equation}
       \mathcal{M}_{z}(\theta) = \left(\begin{matrix}
				\cos\theta & -\sin\theta & \\
				\sin\theta & \cos\theta &\\
                & & 1\\
			\end{matrix}\right) \quad\quad \mathcal{M}_{x}(\zeta) =\left(\begin{matrix}
                1& & \\
				&\cos\zeta & -\sin\zeta \\
				&\sin\zeta & \cos\zeta\\
			\end{matrix}\right).
   \end{equation}
   
   \noindent Next, we have the evolution of $\mathbf{b}$ via

   \begin{align}
			\left(\begin{matrix}
				b_{x} \\
				b_{y} \\
                b_{z} \\
			\end{matrix}\right)=\mathcal{M}_{z}^{-1}(\theta)\mathcal{M}_{x}^{-1}(\zeta)\mathcal{M}_{z}(\Omega t)\mathcal{M}_{x}(\zeta)\mathcal{M}_{z}(\theta)
			\left(\begin{matrix}
				\delta/2 \\
				0 \\
                1-\delta^{2}/8 \\
			\end{matrix}\right)
			\label{eq:b_rot},
	\end{align}
   
   \noindent where $t=\Omega(z-z_{0})/c$. Expanding the expressions to the lowest order of $\Omega t$, the components of $\mathbf{b}$ and $\mathbf{q}$ used in Eq.~\ref{eq:mode_coupling} and Eq.~\ref{eq:R_para} can then be written (for the case of $\alpha>\zeta$) as

   \begin{equation}
			\left\{
			\begin{aligned}
				b_{x}(z)&=\dfrac{\delta}{2}-\Omega \dfrac{z-z_{0}}{c}\sin\zeta \dfrac{\sqrt{\delta^{2}-(\alpha-\zeta)^{2}}}{\delta}\mathrm{sign}((\mathbf{\Omega}\times\mathbf{\mu})\cdot\mathbf{k}),\\
				b_{y}(z)&=- \Omega \dfrac{z-z_{0}}{c}\sin\zeta \dfrac{\alpha-\zeta}{\delta},\\
                b_{z}(z)&=\sqrt{1-b_{x}^2-b_{y}^2},
			\end{aligned}
			\right.
			\label{eq:bxby}
	\end{equation}

    \begin{equation}
			\left\{
			\begin{aligned}
				q_{x}(z)&=\dfrac{\Omega z}{c}\sin\zeta\dfrac{\alpha-\zeta}{\delta},\\
				q_{y}(z)&=-\dfrac{\Omega z}{c}\sin\zeta \dfrac{\sqrt{\delta^{2}-(\alpha-\zeta)^{2}}}{\delta}\mathrm{sign}((\mathbf{\Omega}\times\mathbf{\mu})\cdot\mathbf{k}),\\
                q_{z}(z)&=0.
			\end{aligned}
			\right.
			\label{eq:q_vec}
	\end{equation}

    Given the geometry parameters $(\alpha,\zeta,\delta)$, plasma parameters $(\kappa,\gamma)$, rotation parameters $(\Omega,\dot{\Omega})$, and integration limits $(z_{0},z_{1})$, we can calculate $a_{x,y}$ at a frequency of $\nu=\omega/2\pi$. Thus, we can calculate Stokes parameters $(I,Q,U,V)$, which can then be compared with the observational data. 

\section{Data reduction}\label{sec:data}

Among the parameters noted in the previous paragraph, $(\Omega,\dot{\Omega})$ can be determined once we have access to data from pulsars that are well timed. A number of other parameters could be determined by comparing the observed polarization with modeled polarization. Since mode coupling makes the polarization position angle (PA) deviate from the rotating vector model (RVM, \citealp{RVM}), while producing circular polarization~\citep[e.g.,][]{petrova2003, bp2012}, a full fitting should include both PA and $V$. 

To reduce parameters, we divided the fitting procedure into two steps. First, we derived the geometry parameters $(\alpha,\zeta,\delta)$ from the RVM fitting of integrated pulse profiles. Second, we fixed the geometry parameters and derived the plasma parameters by fitting the circular polarization. In this work, we chose data samples from pulsars whose PA curves were well described by the RVM and we used their single pulses for study. We note that the propagation process occurs in one single pulse and circularly polarized signals with opposite handedness undergo mutual cancellation when translating them into averaged profiles.

Employing the catalog in~\cite{fast682}, we chose three pulsars that have FAST released data\footnote{Their further information could be found on https://fast.bao.ac.cn/observation\_log/search.}: B0301$+$19, J0631$+$1036, and B0656$+$14 (monogem). Their basic parameters are listed in Table~\ref{table:pulsars}, and their locations on the $P$ - $\dot{P}$ diagram are shown in Figure~\ref{fig:ppdot} in Appendix~\ref{sec:RVM}. The data processing (including folding, RFI mitigation, saturation mitigation, calibration, and timing) was performed with the software packages \textsc{dspsr} \citep{2011PASA...28....1V}, \textsc{psrchive} \citep{2004PASA...21..302H}, and \textsc{tempo2} \citep{2006MNRAS.369..655H}. We follow the PSR/IEEE convention for the definition of Stokes parameters \citep{2010PASA...27..104V}. We used a Bayesian method to fit for the rotation measure (RM) of Faraday rotation \citep{2020Natur.586..693L} from the calibrated data. For the linear polarization intensity, $L$, total polarization intensity, $P$, and ellipticity angles, we corrected their biases using the formulae in \cite{cao2025}. The RVM fitting results are also included in Table~\ref{table:pulsars} and the fitting details are given in Appendix~\ref{sec:RVM}.

To improve the signal-to-noise ratio (S/N), we merged the 4096 (or 1024) frequency channels of the data into eight sub-bands. We calculated $V/P$ and $\sigma_{V/P}=\sqrt{P^{2}\sigma_{V}^{2}+V^{2}\sigma_{P}^{2}}/P^{2}$ for the fitting. The pulses used in our study were chosen to satisfy that the minimum $\sigma_{V/P}$  in the pulse profile is smaller than $0.2$. The number of pulses chosen for three pulsars are given in Table~\ref{table:pulsars}. For each pulse, we chose three pulse longitudes $\phi_{1,2,3}$, where $\sigma_{V/P}$s are smallest, to get their $V/P$-$\nu$ spectra. Finally, we had $3\times 8=24$ data points per pulse for the fitting procedure. 

\section{Computational Settings}\label{sec:computational}

If we denote the mode coupling model as $\mathcal{M}$, parameters as $\Theta$, and data as $\mathcal{D}$, then we can use the Bayes theorem to compute the posterior probability distribution function (PDF) via

\begin{equation}
    P(\mathcal{M},\Theta|\mathcal{D})\propto P(\mathcal{M},\Theta)P(\mathcal{D}|\mathcal{M},\Theta),
    \label{eq:Bayes}
\end{equation}

\noindent where $P(\mathcal{M},\Theta)$ is the priorior PDF and $P(\mathcal{D}|\mathcal{M},\Theta)$ is the likelihood function. The likelihood function could be written as

\begin{equation}
    P(\mathcal{D}|\mathcal{M},\Theta) \propto \exp\left(-\dfrac{1}{2}\Sigma_{\phi,\nu}\dfrac{[(V/P)_{\mathcal{D}}(\phi,\nu)-(V/P)_{\mathcal{M},\Theta}(\phi,\nu)]^{2}}{\sigma_{V/P}^{2}(\phi,\nu)}\right).
    \label{eq:likelihood}
\end{equation}

Our $P(\mathcal{D}|\mathcal{M},\Theta)$ contains the numerical solution of Eqs.~\ref{eq:mode_coupling}, where we used package \texttt{odes}~\citep{odes} for solving. We used the \texttt{python} package \texttt{emcee}~\citep{emcee} (version: 3.1.6) for Markov Chain Monte Carlo (MCMC) computations.

We set five parameters in one fitting: $\kappa$, $\gamma/\kappa^{1/3}$, and $z_{0,(1,2,3)}$ for three chosen longitudes. Since the three chosen longitudes tend to be close to each other, we set $\kappa$ and $\gamma$ same for them (i.e., independently of longitudes). We did not use $\gamma$ directly as parameter because, in practice, we find that $\kappa$ and $\gamma$ are sometimes highly correlated. The priorior PDFs of the five parameters are chosen to be $\lg(\kappa)\sim\mathcal{U}(0,7)$, $\lg(\gamma)\sim\mathcal{U}(0.5,3)$, and $\lg(z_{0,(1,2,3)}/R_{\mathrm{NS}})\sim\mathcal{U}(1,3)$, where $\lg$ is the base-10 logarithmic function and $\mathcal{U}(a,b)$ denotes a uniform distribution in the range ($a$, $b$). In addition, $R_{\mathrm{NS}}$ is the neutron star radius and we set $R_{\mathrm{NS}}$ to 10 km. The details of our settings in solving the Bayesian analysis and Eqs.~\ref{eq:mode_coupling} are given below.

\subsection{The priorior PDFs}

The choice of the priorior PDFs' ranges firstly satisfies $\gamma^{2}\gg 1$ and $\kappa\ge1$. We set wide ranges for $\gamma$ and $\kappa$ to explore the parameter space. Since the validity of Eqs.~\ref{eq:mode_coupling} requires high frequency waves, we put an extra constraint on the priorior PDFs via
\begin{equation}
    R(\kappa, \gamma, z_{0})c/\omega_{\mathrm{min}} < 0.1
    \label{eq:prior_constrain},
\end{equation}
\noindent where $\omega_{\mathrm{min}} = 2\pi\cdot 1000$ MHz. We set 128 Markov chains for running the algorithm. To generate the required priorior PDFs, we used \texttt{random} of \texttt{numpy} to generate random values within the ranges we set. If the generated random values satisfy Eq.~\ref{eq:prior_constrain}, they are sampled.

\subsection{From the pulse longitude $\phi$ to $\delta$}

With $\phi_{0}$ obtained from the RVM fitting, the angle between wave vector and magnetic axis, $\delta$, could be derived for each longitude, $\phi$,

\begin{equation}
    \cos\delta = \hat{k}\cdot\hat{\mu}=\sin\alpha\sin\zeta\cos(\phi-\phi_{0})+\cos\alpha\cos\zeta
    \label{eq:delta}.
\end{equation}

\subsection{Considering $z_{0}$ and $z_{1}$}

In \cite{petrova2003}, there were two additional assumptions employed when solving the mode coupling equation: The initial values were set at $z\sim0$, or $z\ll z_{p}$. Furthermore, the final states refer to $z\rightarrow+\infty$. In practice, we have found that the solution of Eqs.~\ref{eq:mode_coupling} is sensitive to $z_{0}$; therefore, we set $z_{0}$ as a parameter to be fitted for each pulse longitude, while $z_{1}$ is the altitude where the magnetic field is too weak to sustain the fourth assumption in Eqs.~\ref{eq:mode_coupling} and could be estimated with the following equation (e.g., \citealp{petrova2001}):

\begin{equation}
    \gamma\omega(1-v_{b}(z_{1})b_{z}(z_{1})/c)=0.1\cdot \dfrac{eB(z_{1})}{m_{e}c}=0.1\cdot \dfrac{eB_{\mathrm{surf}}R_{\mathrm{NS}}^{3}}{z_{1}^{3}m_{e}c}
    \label{eq:z1},
\end{equation}

\noindent where $B_{\mathrm{surf}}$ is the surface magnetic field strength, which could be derived from $P$ and $\dot{P}$ by assuming that magnetic dipole radiation dominates the pulsar spin-down~\citep[e.g.,][]{handbook}. Before running MCMC, we calculated $z_{1}$ using Eq.~\ref{eq:z1}, of all $(\kappa, \gamma/\kappa^{1/3},z_{0},\phi)$ in the posterior PDFs. We chose the smallest $z_{1}$ (shown in Table~\ref{table:pulsars}) as the ending point of all integrations during MCMC. Since $z_{1}$ actually depends on the pulse longitude, $\phi$, we only give an approximate value in Table~\ref{table:pulsars}.

\subsection{Initial polarization state at $z_{0}$}
The polarization state at $z_{0}$ refers to the pure linear polarization as a decoupled solution (Eq. 3.3 in \citealp{lp1998}) of Eqs.~\ref{eq:mode_coupling}. We chose the initial polarization to be O mode and expressed as

\begin{equation}
    \dfrac{a_{x}(z_{0})}{a_{y}(z_{0})}=\dfrac{b_{x}(z_{0})+q_{y}(z_{0})}{b_{y}(z_{0})-q_{x}(z_{0})}.
\end{equation}

\noindent In practice, we fixed $a_{x}(z_{0})=1$. 

\subsection{\texttt{emcee} settings}

We used the \texttt{EnsembleSampler} function of \texttt{emcee} for our computation. To achieve a quicker convergence, the update strategy of the coordinates of walkers are set to be a mixture of different moves: 60\% of \texttt{StretchMove}~\citep{emcee}, 20\% of differential evolution Move (\texttt{DEMove}, \citealp{demove}), and 20\% of snoocker move using differential evolution (\texttt{DESnoockerMove}, \citealp{tv2008}). Details on these moves can be found in the emcee documentation \footnote{https://emcee.readthedocs.io/en/stable}. After some tests on the data, we finally set 10000 steps for running and we omitted the first 6000 steps as burn-in steps, applicable to all pulses.

\section{Results}\label{sec:results}

Before applying the algorithm to real data, we performed a set of tests on simulated data. More details are given in Appendix~\ref{sec:test_data}.

We applied the algorithm to pulses from the three chosen pulsars. For each pulse, we were able to obtain the maximum-likelihood parameters, $\Theta_{m}$, according to the resulting posterior PDF of Bayesian analysis. The results could be divided into two parts: the theoretical curve with a $\Theta_{m}$ versus data curve and the posterior PDF of parameters. 

\begin{figure*}[h!]
   \centering
    \begin{minipage}{\textwidth}
\centering
\includegraphics[width=16cm]{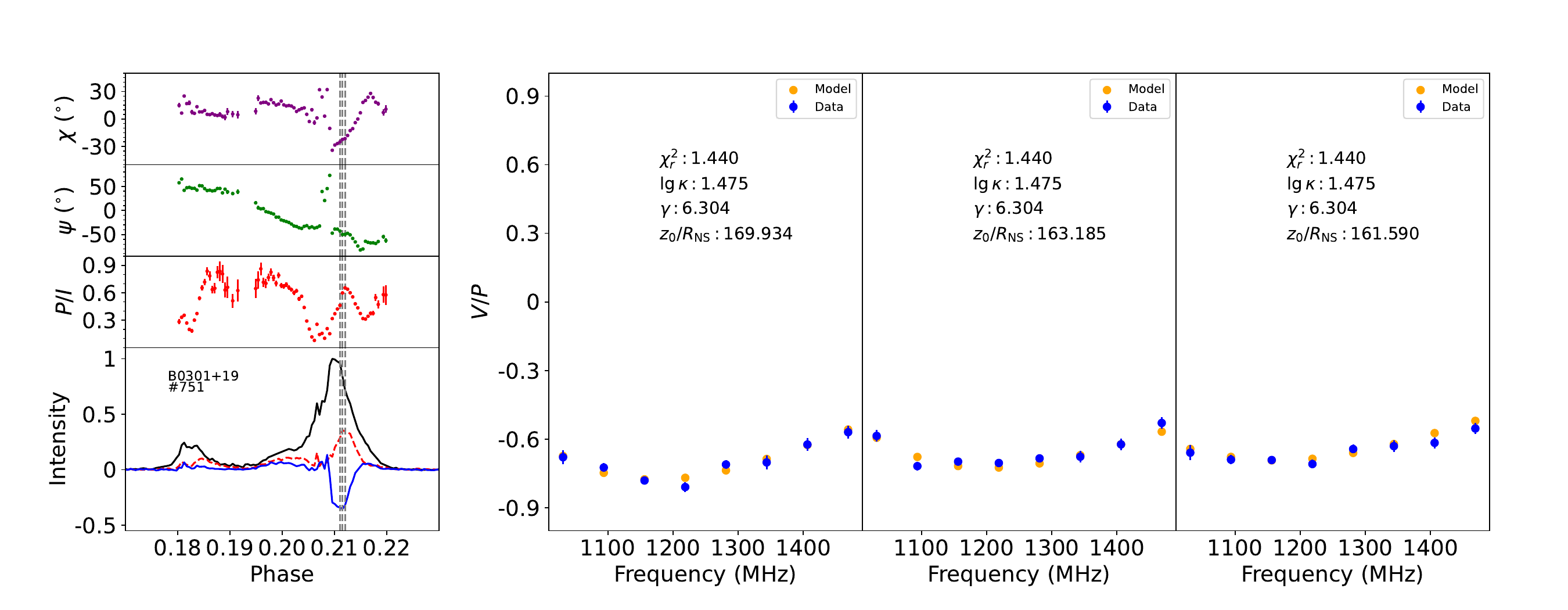}
\raisebox{0cm}{\includegraphics[width=16cm]{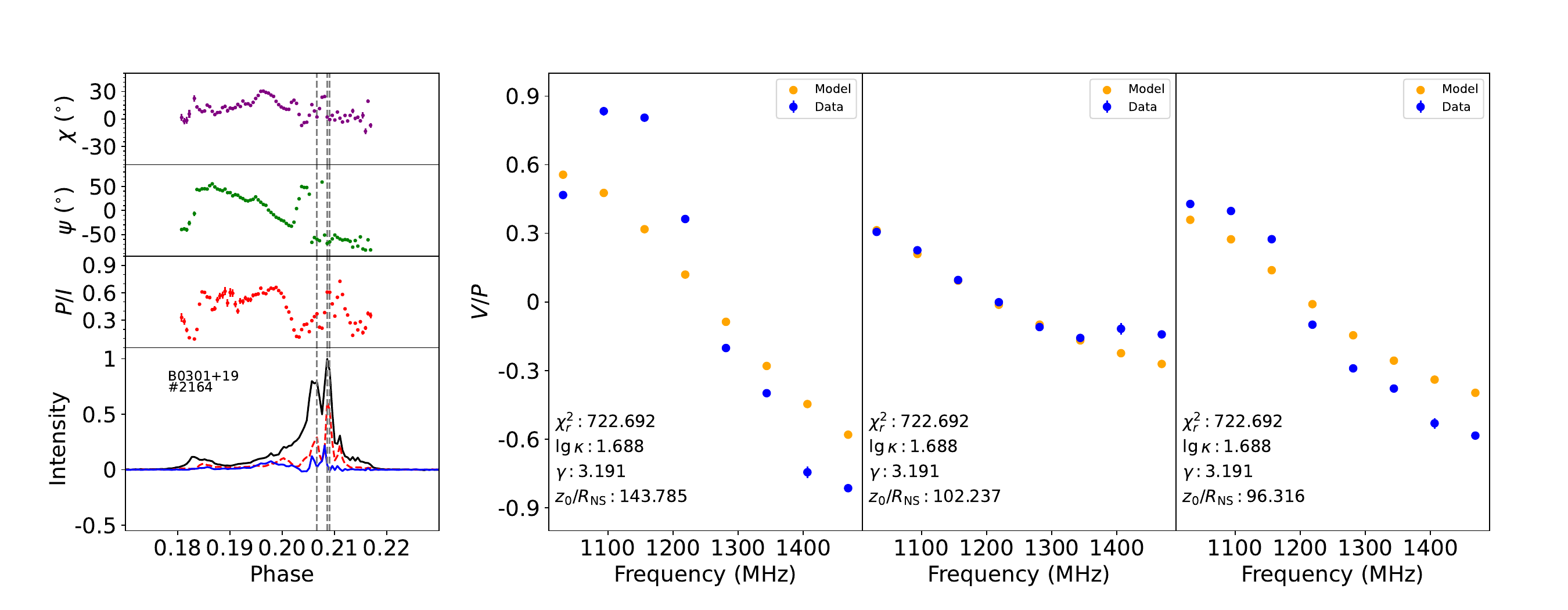}}
\end{minipage}
      \caption{Two pulses (6 $V/P$ - $\nu$ curves) with fitting results. The pulsar name and the pulse number after \# are marked in every figure. For every subfigure, the left panels are polarization profiles of single pulses, and the right plots are $V/P$ - $\nu$ curves of three chosen pulse longitudes (blue dots with errorbars: observation data points; orange dots: modeling points with maximum-likelihood parameters). The maximum-likelihood parameters, and the $\chi^{2}_{r}$s for fitting the three phases are marked out in the right plots. In the Intensity panel: Black line shows the total intensity ($I$); red dashed line shows the linear polarization intensity ($L=\sqrt{Q^{2}+U^{2}}$); blue line shows the circular polarization intensity ($V$). In addition, $I$, $L$, and $V$ are normalized by the maximum total intensity of the respective profiles. The horizontal axis is the pulse phase (0 to 1 in a period). Red dots in the $P/I$ panel with errorbars are the polarization degrees. Green dots in the $\psi$ panel with errorbars are the polarization position angles (PA, $\psi=0.5\arctan(U/Q)$). Magenta dots in the $\chi$ panel with errorbars show the ellipticity angel (EA, $\chi=0.5\arcsin(V/{\sqrt{Q^{2}+U^{2}+V^{2}}})$).}
    \label{fig:well_fitted1}
\end{figure*}

Let us first look at the pulses relatively well-described with the mode coupling model. An example (\#751 of B0301$+$19) is shown in Figure~\ref{fig:well_fitted1}, and more examples are shown in Figure~\ref{fig:more_examples}. The mode coupling model could reproduce the diversity of circular polarization spectra: the circular polarization could be left-handedness (\#683 of B0301$+$19) or right-handedness (\#751 of B0301$+$19) across the frequency band, be approaching 80\% (\#751 of B0301$+$19) or be close to zero (\#35 of J0631$+$1036), and change its handedness when frequency changes (\#1299 of B0301$+$19). The $V/P$ versus $\nu$ curves could also be significantly different between neighbor phases (\#1299 of B0301$+$19), under fixed $\kappa$ and $\gamma$.

A badly fitted pulsation (\#2164 of B0301$+$19) is also shown in Figure~\ref{fig:well_fitted1} (more in Figure~\ref{fig:more_examples}). In those $V/P$ - $\nu$ curves, $V/P$ varies vastly with frequency, but the mode coupling model only reproduces a part of them (e.g., the second plot of \#2164 of B0301$+$19), or even totally fails to reproduce (\#1235 of B0301$+$19). It is worth noting that the chosen phases in these two pulses exhibit PA jumps, which may indicate a more complex mode mixing than pure-mode coupling described by Eqs.~\ref{eq:mode_coupling}. We discuss this further in Section~\ref{sec:discussions}.

\begin{figure*}[h!]
   \centering
    \includegraphics[width=18cm]{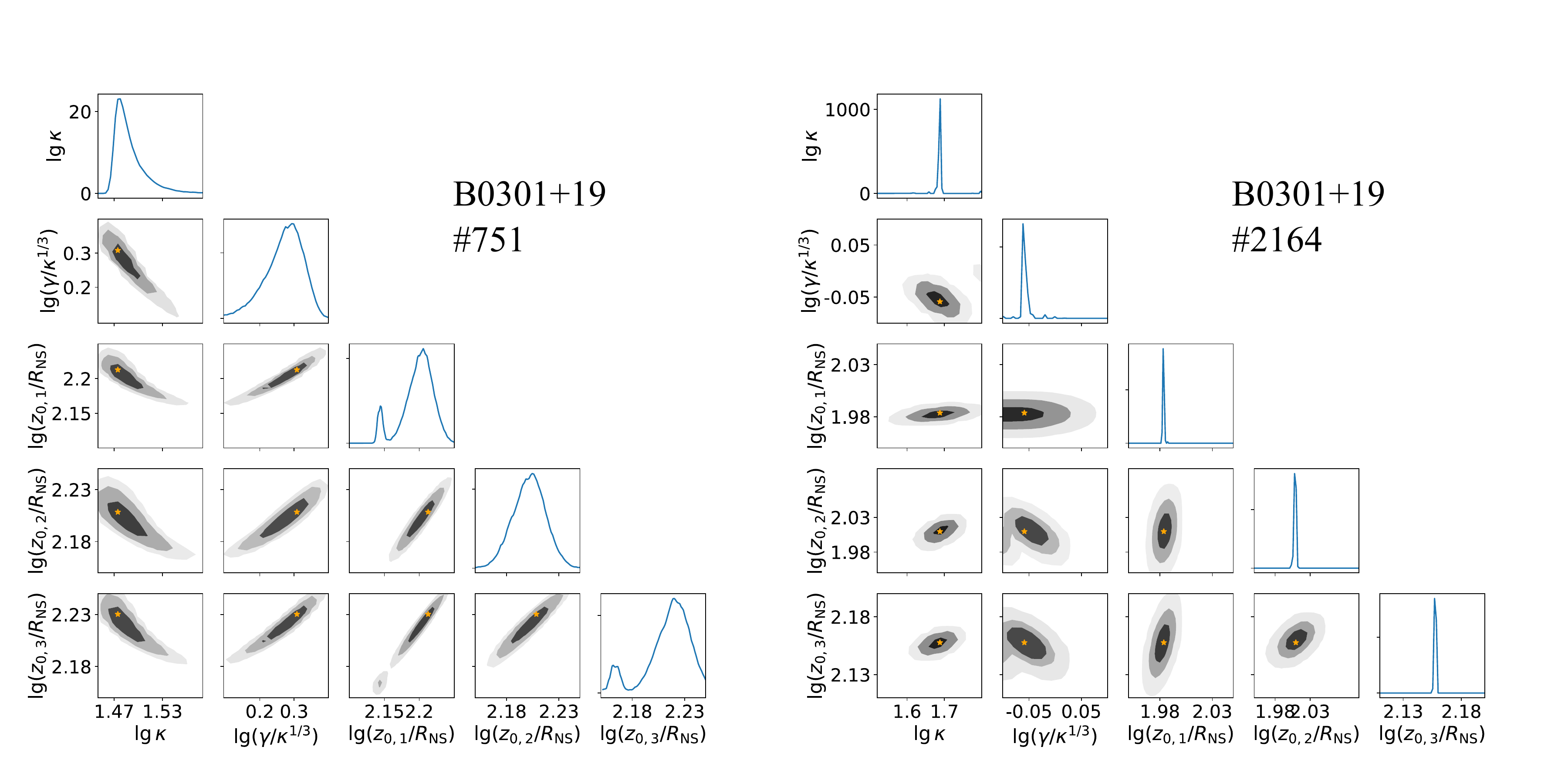}
    \caption{Posterior PDFs of two pulses in Figure~\ref{fig:well_fitted1}. The pulsar names and pulse numbers are marked in each plot. The blue curves are marginal PDFs of parameters and the grey contours are 2D PDFs of each two parameters. The contour profiles mark the credibility regions of probabilities 50\%, 80\%, and 94\%. The orange stars represent the maximum-likelihood parameters.}
    \label{fig:contour_real}
\end{figure*}

Three examples of posterior PDFs are presented in Figure~\ref{fig:contour_real} and Figure~\ref{fig:multi_peaks}. The shapes of the posterior PDFs vary manifestly among different pulses: some are strongly multipeaked (e.g., \#683 of B0301$+$19), while some have much simpler shapes that are closer to single-peaked (e.g., \#751 and \#2164 of B0301$+$19). This is also reflected in the posterior PDFs of simulated data: in Figure~\ref{fig:simulated_003} of Appendix~\ref{sec:test_data}, we see that $z_{3}$ is more tightly constrained than $z_{1}$ and $z_{2}$. We used a clustering algorithm with \texttt{hdbscan}\footnote{https://hdbscan.readthedocs.io/en/latest/} package to pick out three peaks in the posterior PDF of \#683 of B0301$+$19, and to compare the modeling $V/P$ - $\nu$ with the data points. The results are shown in Figure~\ref{fig:multi_peaks} of Appendix~\ref{sec:comparison_app}, indicating that different groups of parameters could lead to similar circular polarization spectra.

To evaluate the goodness of fitting, we calculated the reduced chi-square $\chi^{2}_{r}$ of every pulse,

\begin{equation}
    \chi^{2}_{r} = \dfrac{1}{24-5}\Sigma_{\phi,\nu}\dfrac{[(V/P)_{\mathcal{D}}(\phi,\nu)-(V/P)_{\mathcal{M},\Theta_{m}}(\phi,\nu)]^{2}}{\sigma_{V/P}^{2}(\phi,\nu)}.
\end{equation}

\begin{figure*}[htbp]
\centering
\includegraphics[width=18cm]{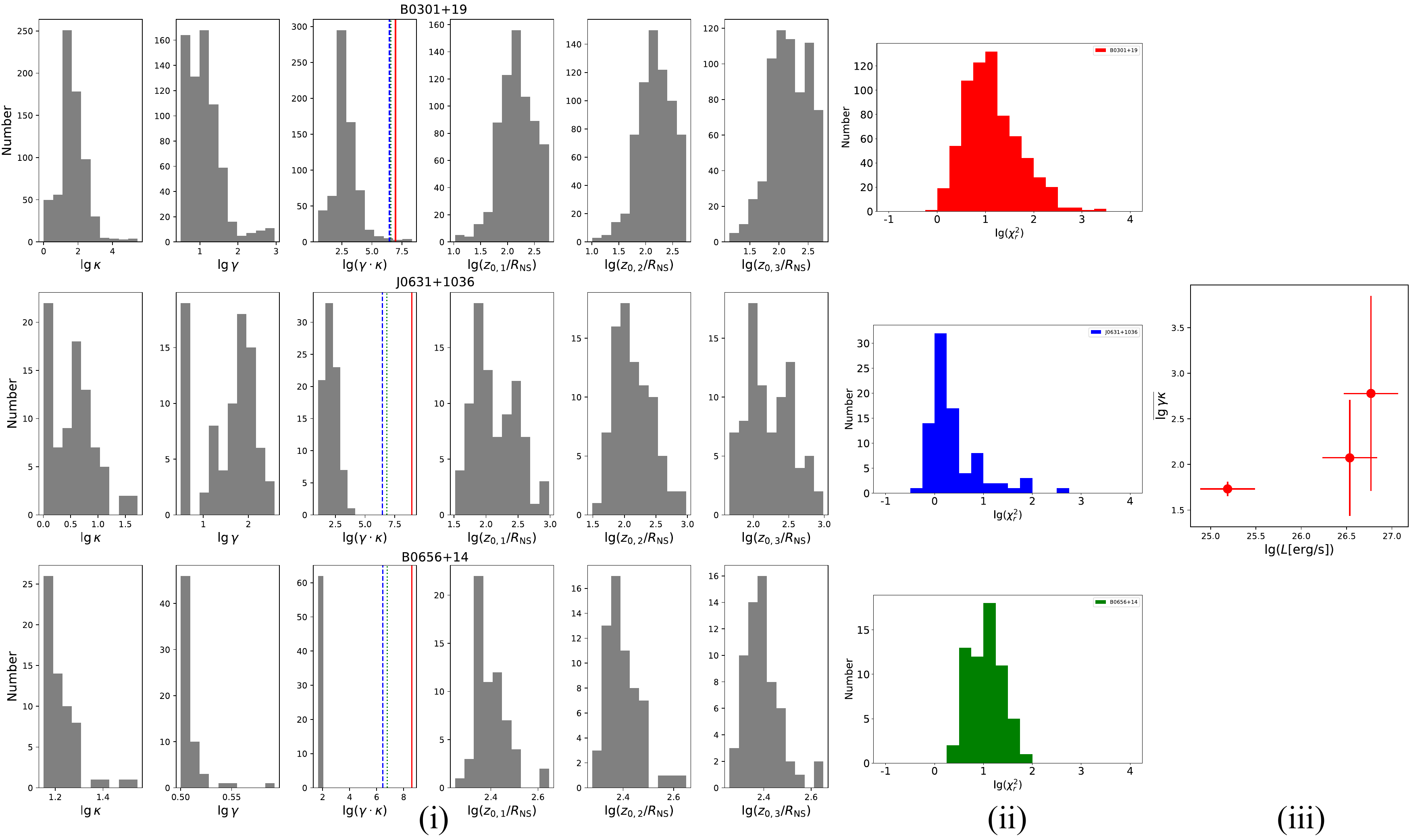}

\caption{(i) Distributions of maximum-likelihood parameters of MCMC results for three pulsars. The third panels of three plots are distributions of multiplicity times Lorentz factor (logarithmic). In each $\lg(\gamma\cdot\kappa)$ distributions: the red vertical line represents the maximum potential drop in polar cap region times the unit charge, $e$, divided by electron rest energy $m_{e}c^{2}$; the blue vertical dashed line is the inner gap potential drop $\times e/(m_{e}c^{2})$ given in \cite{rs1975}; the green vertical dotted line is the charge starvation zone potential drop $\times e/(m_{e}c^{2})$ given in \cite{as1979}. For details, see Section~\ref{sec:kappa_discussion}. (ii) Common logarithmic values of reduced chi-square distributions for the fittings of three pulsars, in the form of histograms. For the horizontal axis, The (-1,4) range is divided into 20 bins. (iii) The correlation between averaged $\kappa\gamma$ and radio luminosity $L$ for three pulsars. For details please refer to Section~\ref{sec:kappa_discussion}.}
\label{fig:chi2_r}
\end{figure*}

An evident conclusion could be drawn from the $\chi^{2}_{r}$ distribution (Figure~\ref{fig:chi2_r}): the mode coupling model is not accurate enough to quantitatively explain the circular polarization of most chosen pulses, with $\lg\chi_{r}^{2}$ significantly deviating from 0.

\begin{figure*}[htbp]
\centering
\includegraphics[width=18cm]{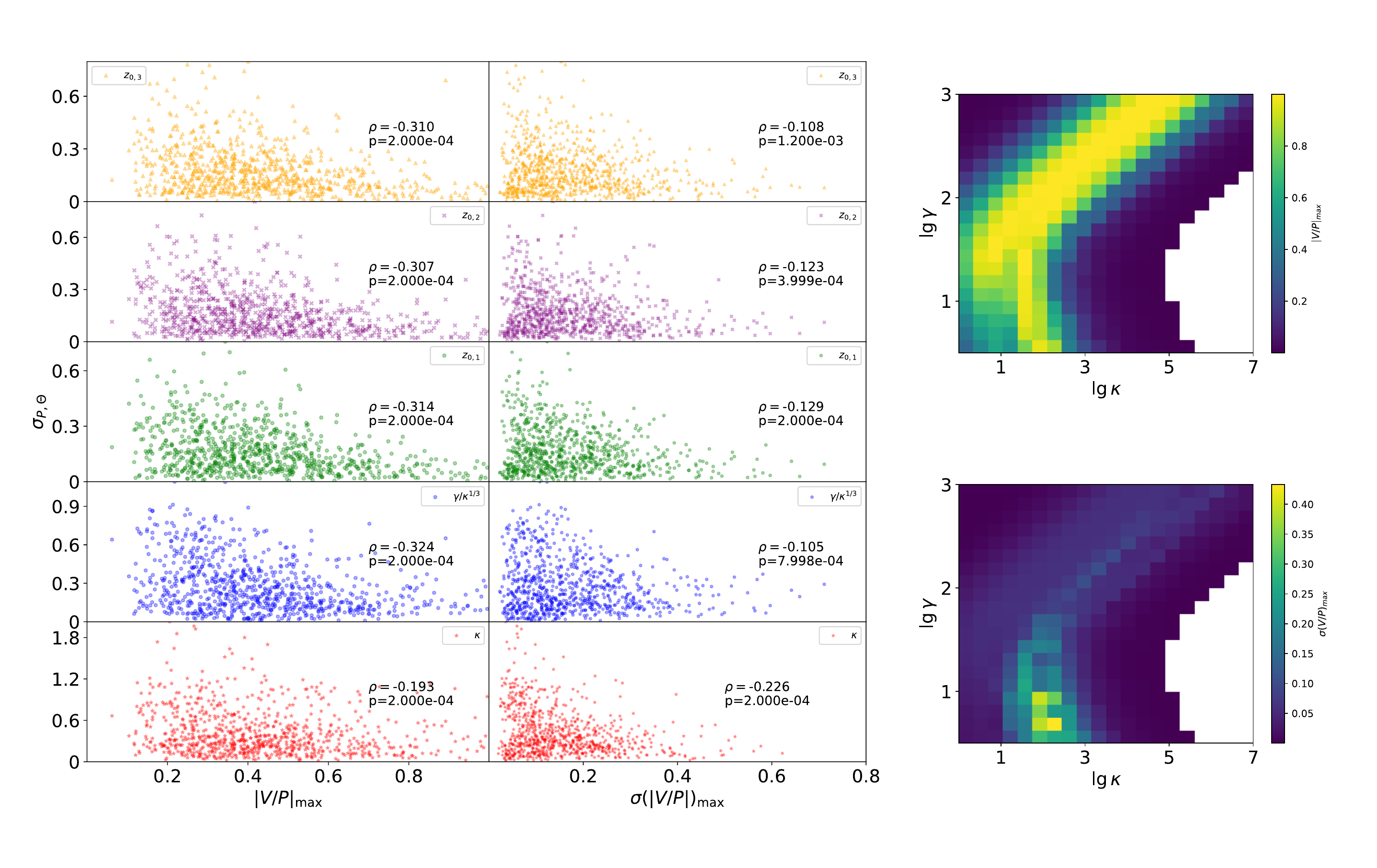}
\caption{Left two plots: Correlations between standard deviations of posterior PDFs of five modeling parameters and the maximum circular polarization fractions in pulses (left), or the maximum standard deviations of circular polarization fraction spectra in pulses (right). The Spearman rank correlation coefficients, $\rho$, and the p-value of permutation tests are marked in all panels. For details, see the main text. Right two plots: Maximum $|V/P|$ and the maximum $\sigma(V/P)$ calculated from the mode coupling model, for pairs of $(\lg\kappa, \lg\gamma)$. The calculation details are given in the text. The blank regions in the plots are the parameter spaces, where $Rc/\omega\ll 1$ is violated.}
\label{fig:corr_sigma}
\end{figure*}

To explore whether the properties of the $V/P$ - $\nu$ curve affect the goodness of parameter constraint, first we drew a correlations between the standard deviations, $\sigma_{P,\Theta}$, of the posterior PDFs of five parameters and the largest absolute $V/P$ values ($|V/P|_{max}$). For each pulse, we also calculated the standard deviations of $V/P$ values on each longitude and obtained the largest one $\sigma(V/P)_{\mathrm{max}}$ out of three longitudes. The correlations between $\sigma_{P,\Theta}$ and $\sigma(V/P)_{\mathrm{max}}$ are also plotted. The results for B0301$+$19 are shown in Figure~\ref{fig:corr_sigma}. A larger circular polarization fraction or a higher varying circular polarization spectrum tend to constrain the parameters more tightly. We quantified the monotonic dependence between $|V/P|_{max}$ (and $\sigma(V/P)_{\mathrm{max}}$) and $\sigma_{P,\Theta}$ using the Spearman rank correlation coefficient, $\rho$. To account for the nonuniform sampling in A, we assessed the statistical significance via a permutation test, in which the $\sigma_{P,\Theta}$ values are randomly shuffled while keeping $|V/P|_{max}$ (and $\sigma(V/P)_{\mathrm{max}}$) fixed. We used a p-value to describe the probability of achieving a more negative $\rho$ in the permutation test. The p-value $p$ is computed as $(k+1)/(N_{\mathrm{perm}}+1)$, where $k$ is the number of permuted samples yielding a correlation coefficient more extreme than the observed one, and $N_{\mathrm{perm}}$ is the number of times of permutation. The $\rho$ and $p$ are marked in all panels of Figure~\ref{fig:corr_sigma}. The observed correlation is found to be significantly more negative than expected under the null hypothesis of no correlation.

The distributions of $\Theta_{m}$s, separated for three pulsars, are shown in Figure~\ref{fig:chi2_r}. The fitting results are different for them. The distribution of $\kappa$ for J0631$+$1036 is peaked at the lower limit $10^{0}$, while the distribution of $\gamma$ for B0656$+$14 is even more concentrated around the lower limit value $10^{0.5}$. Regarding what is shown in Figure~\ref{fig:chi2_r}, the $\chi_{r}^{2}$ distribution of B0656$+$14 generally deviates from $\chi_{r}^{2}=10^{0}$, the model fails to quantitatively describe the chosen pulses of B0656$+$14.

\section{Discussions}\label{sec:discussions}

\subsection{Diversity of single pulse circular polarization}

The highly variable single pulse morphology hints at the variations among the magnetospheric parameters between pulse and pulse, as well as the variety of parameters, leading to the diversity of single pulse circular polarization. The $V/P$ change of both absolute value and handedness could be attributed to the change of the amplitude ratio and phase lag between $a_{x}$ and $a_{y}$. The model fails to quantitatively explain all pulses; this is understandable since the model is greatly simplified.

The goodness of the parameter constraint depends on the $V/P$ - $\nu$ curve properties. To understand this phenomenon from the model side, we calculated the $V/P$ - $\nu$ curves with a $20\times20\times20$ grids of values for $\lg\kappa\in(0,7)$, $\lg\gamma\in(0.5,3)$, and $\lg(z_{0}/R_{NS})\in(0,3)$. Then, for each pair of $(\lg\kappa, \lg\gamma)$, we found the maximum $|V/P|$ and the maximum $\sigma(V/P)$, and plotted them: the result is shown in the right two plots of Figure~\ref{fig:corr_sigma}. We can conclude that larger $|V/P|_{\max}$ and larger $\sigma(V/P)_{\max}$ have less of a parameter degeneracy than the smaller ones. They require the amplitude ratio and phase lag between $a_{x}$ and $a_{y}$ to simultaneously be in an appropriate range.

\subsection{The effect of orthogonal modes mixture on $V/P$ - $\nu$ curves}

As already pointed out in Section~\ref{sec:results}, the two badly fitted cases in Figure~\ref{fig:well_fitted1} and Figure~\ref{fig:more_examples} are associated with 90$^\circ$ PA jumps. The PA jumps indicate the incoherent (or partially coherent) superposition of orthogonal polarization modes (OPMs)~\citep[e.g.,][]{cr1979,karastergiou2005,oswald2023}. In our computation, we assumed that before propagating into the limiting polarization region, the wave is a pure O-mode. This assumption may break down at PA-jumping longitudes for two plausible reasons: (1) the amplitude ratio of OPMs is expected to be frequency-dependent~\citep{petrova2001} and (2) the partially coherent superposition of OPMs itself also induces circular polarization components~\citep{dyks2017,oswald2023,cao2025}; in addition, the frequency dependence of $V/P$ could be more complex than what could be fitted by the simple mode coupling model we adopted in this work. This qualitatively explains the large $\chi_{r}^{2}$ of the fitting of two pulses in Figure~\ref{fig:well_fitted1}.

\subsection{The physical meaning of $z_{0}$}
Here, $z_{0}$ is the point where mode coupling becomes important. From this aspect, $z_{0}$ seems to be the limiting polarization radius defined by Eq.~\ref{eq:limiting_pol_psr}. We note that Eq.~\ref{eq:limiting_pol_psr} is only an order-of-magnitude estimate and does not allow for an exact determination of $z_{0}$. A more detailed modeling of pulsar magnetosphere may be needed for further discussions of $z_{0}$.

\subsection{Implications of derived plasma parameters on pulsar theories}\label{sec:kappa_discussion}

The multiplicity $\kappa$ and Lorentz factor $\gamma$ values of the three pulsars are different. In particular, $\kappa$ is around $10^{0}$ to $10^{2}$ and $\gamma$ is around $10^{0.5}$ to $10^{2}$. Their $\kappa$ values are generally in accordance with former simulations and theories of the pair cascade in the pulsar magnetospheres~\cite{dh1982, gi1985}, well below the maximum multiplicity estimated by former modelings, usually around $10^{5}$~\citep{bgibook,th2015,th2019}. In addition, \cite{petrova2003} reported $\kappa\sim 1$ - $100$ for PSR~B0329$+$54, which is close to our results. While \cite{galishnikova2020} gives $\kappa\sim 10^{3}$ for PSR~J1906$+$0746. A plausible explanation for the relatively small multiplicity of some pulsars is that the radiation we measured escapes through low-density ``holes'' in the highly nonuniform plasma distributions created by pair cascades, since the nonuniformity is generally suggested in simulations~\citep[e.g.,][]{ta2013, pts2020, benacek2024}.

The question of why our Bayesian analysis does not give larger $\kappa$ and $\gamma$ values might also be related to the energy supplement to the pair cascade. Overall, $\kappa n_{\mathrm{GJ}}\gamma m_{e}c^{2}$ could be regarded as the energy density of the secondary particles and should be restricted by the acceleration processes above the pulsar polar cap. Assuming that the maximum acceleration potential drop is $\Phi$ and that the initial particle number density is $n_{\mathrm{GJ}}$, we would expect to have $\kappa n_{\mathrm{GJ}}\gamma m_{e}c^{2}<e\Phi n_{\mathrm{GJ}}$, i.e., $\kappa\gamma<e\Phi/m_{e}c^{2}$. We used three values of $\Phi$ for testing the inequality. The first one is the potential difference between the magnetic axis and the polar cap boundary~\citep[e.g.,][]{beskin2018}:
\begin{equation}
    \Phi_{\mathrm{PC}} = \dfrac{1}{2} (\Omega R_{\mathrm{NS}}/c)^2  R_{\mathrm{NS}}  B_\mathrm{surf}
    \label{eq:phi_pc}.
\end{equation}
\noindent The second one is the inner gap potential drop given by \cite{rs1975}:
\begin{equation}
    \Phi_{\mathrm{RS}} \approx 1.6\times10^{12} \left(\dfrac{B_{surf}}{10^{12}\mathrm{G}}\right)^{-1/7} \left(\dfrac{2\pi/\Omega}{1\mathrm{s}}\right)^{-1/7}\mathrm{V}
    \label{eq:phi_rs}.
\end{equation}
\noindent The third one is the charge starvation zone potential drop given by \cite{as1979}:
\begin{equation}
    \Phi_{\mathrm{RS}} \approx 5\times10^{12} \left(\dfrac{2\pi/\Omega}{0.1}\right)^{-3/8}\mathrm{V}
    \label{eq:phi_as}.
\end{equation}
The $\lg(e\Phi/m_{e}c^{2})$ values are plotted in Figure~\ref{fig:chi2_r}. We also compared the power carried by plasma particles, $P_{e}$, with the rotating energy loss rate, $\dot{E}$, of pulsars, assuming for the purposes of this study that $\dot{E}$ is equal to dipole energy loss $B_{\mathrm{surf}}^{2}\Omega^{4}R_{\mathrm{NS}}^{6}/6c^{3}$ via

\begin{equation}
    \dfrac{P_{e}}{\dot{E}} = \dfrac{\kappa n_{\mathrm{GJ}}(z=R_{\mathrm{NS}})\gamma m_{e}c^{2}\pi R_{\mathrm{PC}}^{2}c}{-I_{*}\Omega\dot{\Omega}} = \dfrac{3\kappa\gamma m_{e}c^{4}}{eB_{\mathrm{surf}}\Omega^{2}R_{\mathrm{NS}}^{3}}\sim\dfrac{\kappa\gamma}{\Phi_{\mathrm{PC}}e/m_{e}c^{2}},
    \label{eq:ratio}
\end{equation}

where $R_{\mathrm{PC}}=R_{\mathrm{NS}}\sqrt{\Omega R_{\mathrm{NS}}/c}$ is the polar cap radius and $I_{*}$ is the neutron star moment of inertia. Most derived $\kappa\gamma$ values are below those $\lg(e\Phi/m_{e}c^{2})$ values; therefore, the $P_{e}/\dot{E}$ ratio is small. This indicates that acceleration in the inner magnetosphere ($z\lesssim 1000 R_{\mathrm{NS}}$) is far less efficient than the acceleration taking place in the outer magnetosphere or beyond the light cylinder.

Since radio emission usually originates from inner magnetosphere, we explored the correlation between $\kappa\gamma$ and radio luminosity. For each pulsar we chose, the radio luminosity could be estimated as
\begin{equation}
    L\approx (S_{1400}/W_{50})\Delta\nu D^{2}2\pi (1-\cos{\pi W_{50}})\approx S_{1400}\Delta\nu D^{2}\pi^{3}W_{50},
\end{equation}
\noindent where $S_{1400}$ is the mean flux density at 1400~MHz and $\Delta\nu=500$ MHz is observation bandwidth, while $D$ is the distance of pulsar from us and $W_{50}$ is the full width at half maximum (FWHM) of the pulse profile in the form of fraction (e.g., 0.05 = 5\% of the period). We obtained $S_{1400}$ and $D$ from the \textsc{psrcat} website\footnote{https://www.atnf.csiro.au/research/pulsar/psrcat/}~\citep{psrcat}, and set $W_{50}=0.05$. The correlation between $\kappa\gamma$ and $L$ is plotted in Figure~\ref{fig:chi2_r}. The vertical errorbar was calculated from the inferred $\kappa\gamma$ distribution in Figure~\ref{fig:chi2_r}, while the horizontal errorbar is set as $\lg 2$. There seems to be a positive correlation between $\kappa\gamma$ and $L$. This tendency is reasonable because radio emission should originate from the particles we measured, $\kappa\gamma$; namely, the inner magnetospheric plasma particles.

Our algorithm could potentially be applied to more pulsars. However, since there might be other various mechanisms contributing to the pulsar circular polarization, more detailed modeling should be carried out. Finally, we note that B0656$+$14 has been observed to have extended high-energy emission (pulsar wind nebula and pulsar halo, \citealp{birzan2016, abeysekara2017}). Modeling the radiation spectra of pulsar wind nebulae can also constrain the multiplicity~\citep{dejager2007, spencer2025}. Therefore, the multiplicity of B0656$+$14 may be also derived from high-energy observations and be compared with the result of this study in the future.

\section{Conclusions}\label{sec:conclusions}
We performed a quantitative analysis of the circular polarization of single pulses observed by FAST, based on wave-mode coupling equations in the magnetospheres of pulsars, put forward by \cite{lp1998} in the limiting polarization of a pulsar magnetosphere. Furthermore, we explored the details of the model properties. We found that the model could reproduce the diversity of single pulse circular polarization and could give some constraint to the plasma parameters, multiplicity, $\kappa$, and Lorentz factor, $\gamma$. The goodness of fit and the goodness of constraint are all related to the exact shapes of circular polarization spectra curves. Our derived plasma parameters indicate that the inner magnetospheric flow power responsible for coherent radio emission is only a very small fraction of the pulsar spin-down energy loss and the power seems to be positively correlated with spin-down rate.

Over the past fifty years, various models have been developed for explaining pulsar radio emission properties. However, quantitative studies of pulsar radio emission signals are still insufficient. Now that FAST, MeerKAT, and other facilities are producing high-quality radio pulsar data, it is now imperative to test previously established theories with high-quality data to gain a more comprehensive understanding of pulsar magnetospheres.

\begin{acknowledgements}
We thank the Principal Investigators of the released data we use: Mao Yuan and Shijun Dang. We gratefully thank V. S. Beskin, Arsenii Istomin, Fedor Kniazev, and all members of the pulsar group at Peking University for discussions. SSC gratefully thanks Ziming Wang, Yiming Dong and Zezhong Xu for useful discussions on Bayesian analysis. SSC also thanks Yuanhong Qu for encouragement, thanks Chengjun Xia and Qiming Yan for inspiring him to do Bayesian analysis, and thank Heng Xu for suggestions on handling observation data saturation. An anonymous reviewer is acknowledged for providing us with many suggestions. JJC acknowledges financial support from the European Research Council (ERC) starting grant ``COMPACT'' (Grant agreement number 101078094). WYW acknowledges financial support from NSFC No.12403058. All data used in this work is from the FAST (Five-hundred-meter Aperture Spherical radio Telescope) (https://cstr.cn/31116.02.FAST). FAST is a Chinese national mega-science facility, operated by National Astronomical Observatories, Chinese Academy of Sciences. This work is supported by the National SKA Program of China (2020SKA0120100), the National Natural Science Foundation of China (Nos. 12003047 and 12133003), and the Strategic Priority Research Program of the Chinese Academy of Sciences (No. XDB0550300).
\end{acknowledgements}

\begin{appendix}
\section{Information of 3 chosen pulsars, and RVM fitting}\label{sec:RVM}

\begin{figure}[h!]
   \centering
    \includegraphics[width=9cm]{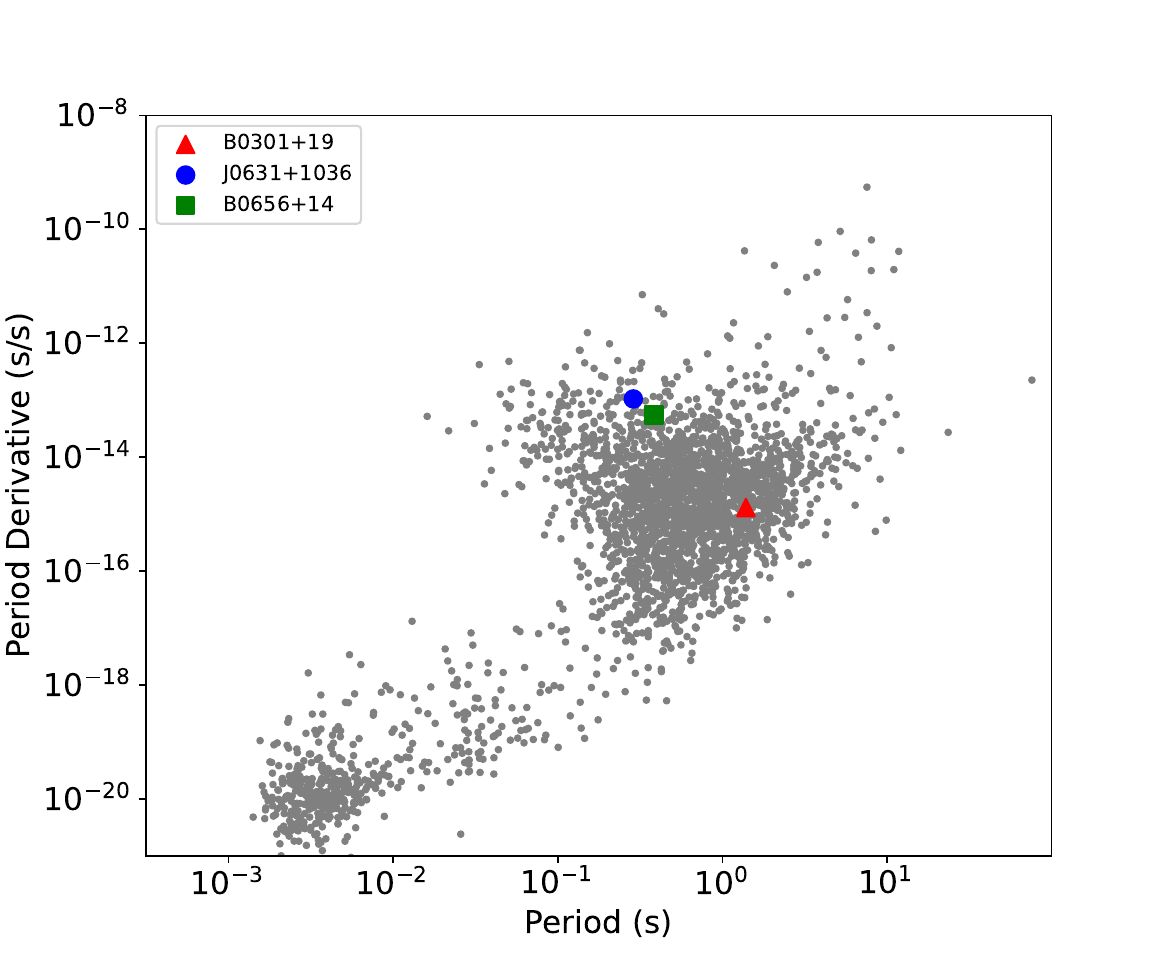}
      \caption{$P$ - $\dot{P}$ diagram. The grey dots are from the \textsc{psrcat}~\citep{psrcat} website (https://www.atnf.csiro.au/research/pulsar/psrcat/). The red triangle, blue dot, and green square mark the three pulsars chosen for our study.}
         \label{fig:ppdot}
\end{figure}

Three selected pulsars are B0301$+$19, J0631$+$1036, and B0656$+$14 (monogem). Their basic parameters are listed in Table~\ref{table:pulsars} and their locations on the $P$ - $\dot{P}$ diagram are shown in Figure~\ref{fig:ppdot}.
\begin{table}[!h]
    \caption{Parameters of chosen pulsars}              
    \label{table:pulsars}    
    \centering                        
    \resizebox{\columnwidth}{!}{
    \begin{tabular}{l l l l l l l l l l}     
    \hline\hline               
    Name & Period (s) & $\dot{P}$ (s/s) & $N_{\mathrm{channel}}$ & Obs. date (UTC)&$z_{1}/R_{NS}$ & $\alpha (^{\circ})$ & $\zeta (^{\circ})$ & $\phi_{0}$ & $N_{\mathrm{p},\mathrm{chosen}}$\\         
    \hline                      
    B0301$+$19 & 1.39 & 1.30$\times10^{-15}$ & 4096 & 20210912 &$\sim 560$ & 52.92& 50.23& 0.199 & 679\\    
    J0631$+$1036 & 0.288 & 1.05$\times10^{-13}$ & 1024& 20230110 & $\sim 1040$ & 177.3 & 176.9 & 0.750 & 85\\
    B0656$+$14 & 0.385 & 5.49$\times10^{-14}$ & 4096 & 20230430 & $\sim1090$ & 177.6& 176.7 & 0.769 & 62\\
    \hline                                  
    \end{tabular}
    }
    \tablefoot{$N_{\mathrm{channel}}$ refers to the number of frequency channels. Obs. date (UTC) indicates the observation data in UTC. $\alpha$ is the magnetic inclination angle, $\zeta$ is the view angle, and $\phi_{0}$ is the phase (0 to 1) of the fiducial plane where the spin axis, magnetic axis and line-of-sight are coplanar. Finally, $N_{\mathrm{p},\mathrm{chosen}}$ is the number of pulses chosen for the Bayesian analysis.}
\end{table}

\begin{figure*}[h!]
   \centering
    \includegraphics[width=15cm]{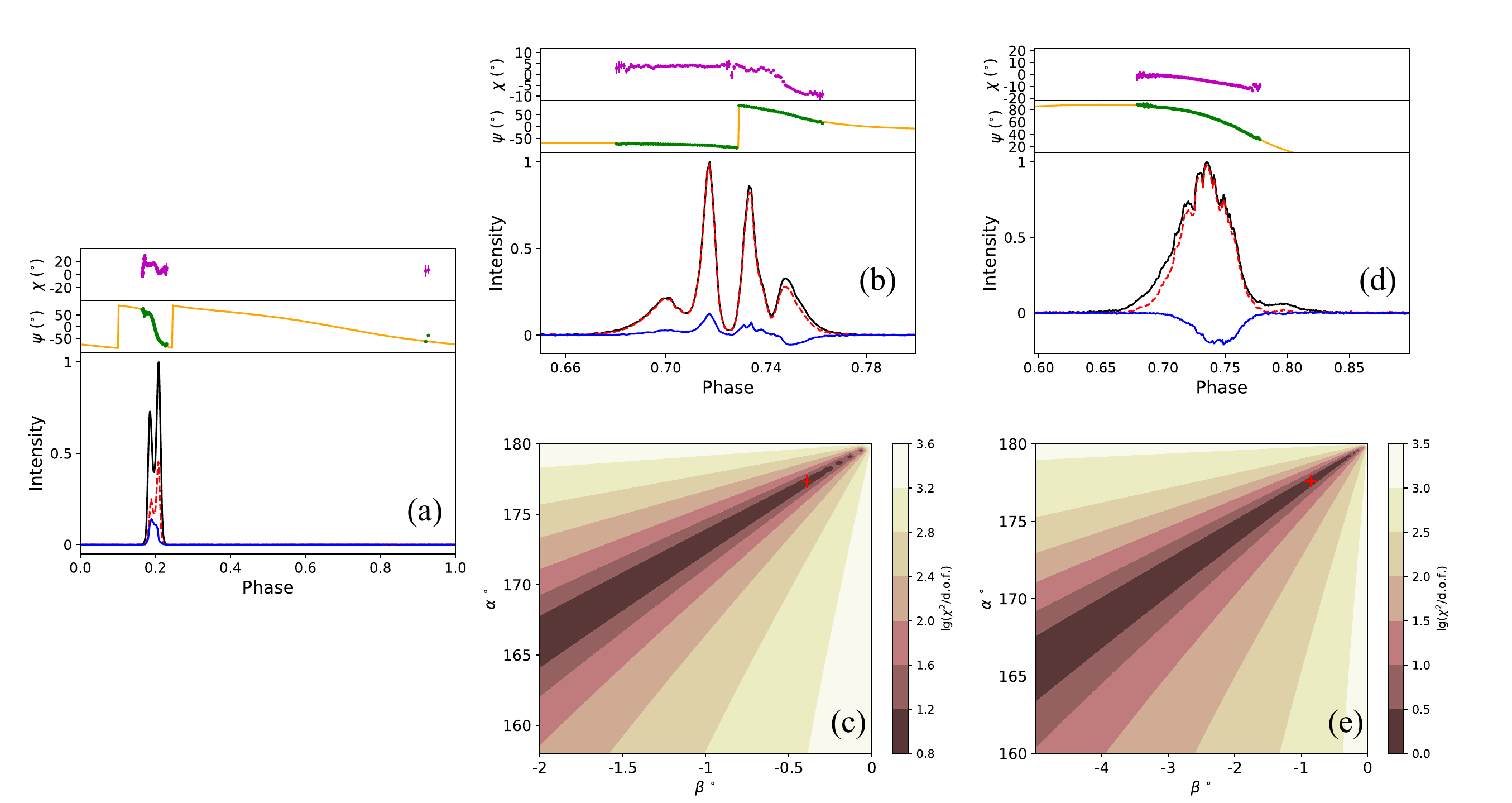}
    \caption{Integrated pulse profiles and RVM fitting results of the three pulsars. (a)(b)(d) are profiles of B0301$+$19, J0631$+$1036, and B0656$+$14, and (c)(e) are $\chi_{r}^{2}$ distributions for $(\alpha,\zeta)$ pairs of J0631$+$1036 and B0656$+$14. The meanings of lines and dots in profiles are same as those in Figure~\ref{fig:well_fitted1}, except for the orange-colored RVM curves of fitted parameters. }
    \label{fig:prof_geo}
\end{figure*}

The RVM fitting algorithm is same as that in \cite{cao2024} and \cite{cao2025}: for each pair of $(\alpha,\zeta)$, we calculate the minimum reduced chi-square ($\chi_{r}^{2}$) by varying $\phi_{0}$ and $\psi_{0}$. For RVM fitting, We chose PA values from pulse longitudes where linear polarization $L$ is significant enough: $L/\sigma_{L}>10$. This basic algorithm works relatively well for J0631$+$1036 and B0656$+$14. Their $\chi_{r}^{2}$ distributions along with integrated profiles are shown in Figure~\ref{fig:prof_geo}.

When applying the algorithm to B0301$+$19, we find that the minimum $\chi^{2}_{r}$ is always larger than 50, even when we chose only highly linearly polarized pulse bins to produce a highly linearly polarized profile and use its PA (the method used in \citealp{mitra2023} and \citealp{johnston2024}). The radiative geometry of B0301$+$19 has been measured in former papers~\citep{ew2001,yuan2023,sun2025}, and their results are not all the same. Although not significant in our chosen observation, B0301$+$19 profile has an interpulse, which is also pointed out in \cite{yuan2023,sun2025}. The interpulse longitudes does not satisfy $L/\sigma_{L}>10$. But to better constrain the geometry, we pick two PA points within the interpulse range, and add them into the highly linearly polarized ($L/I>0.8$ and $L/\sigma_{L}>10$) PA curve. We do not use the algorithm of calculating minimum $\chi^{2}_{r}$ for $(\alpha, \zeta)$, but just directly use least-squares method to fit the PA curve. The $\alpha$, $\zeta$, and $\phi_{0}$ are given in Table~\ref{table:pulsars}, and the pulse profile with RVM is also shown in Figure~\ref{fig:prof_geo}. The fitting result is close to that in~\cite{bcw1991,yuan2023}.

\section{Bayesian analysis of simulated Data}\label{sec:test_data}
To test how the model works and how the parameters will be constrained, we generated some simulated $V/P$ versus $\nu$ curves from the model. We set $\kappa=450$, $\gamma=4.2$, $z_{0,1}=120R_{\mathrm{NS}}$, $z_{0,2}=115R_{\mathrm{NS}}$, and $z_{0,3}=412R_{\mathrm{NS}}$. We then added Gaussian noise to the simulated $V/P$ values, with standard deviation $\sigma$, and do Bayesian analysis on them. We set 20 simulated $V/P$ - $\nu$ curves using the parameters of B0301$+$19, for $\sigma=0.01$ and $\sigma=0.03$ separately.

\begin{figure*}[h!]
   \centering
   \begin{minipage}{\textwidth}
\centering
\includegraphics[width=8cm]{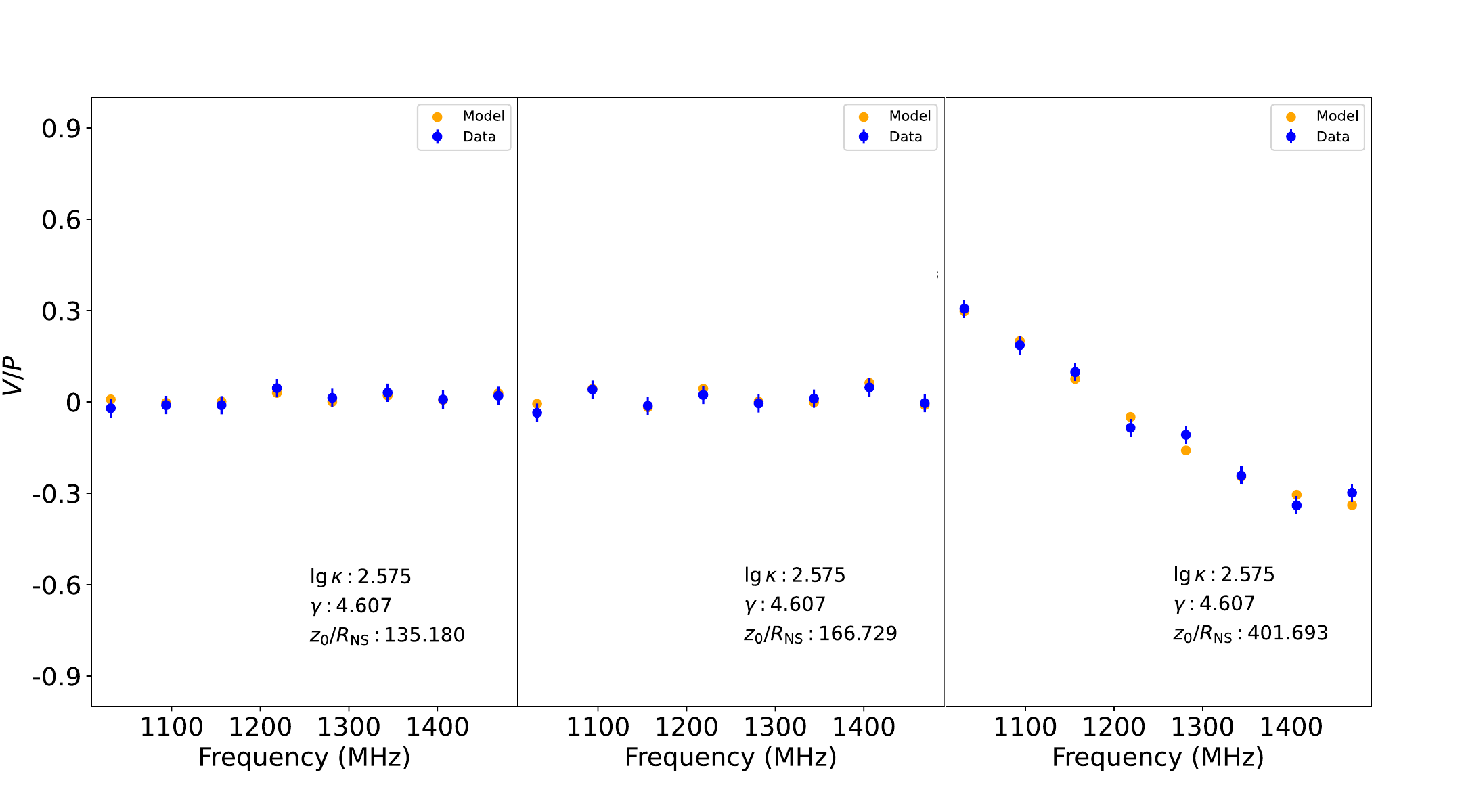}
\raisebox{0cm}{\includegraphics[width=8cm]{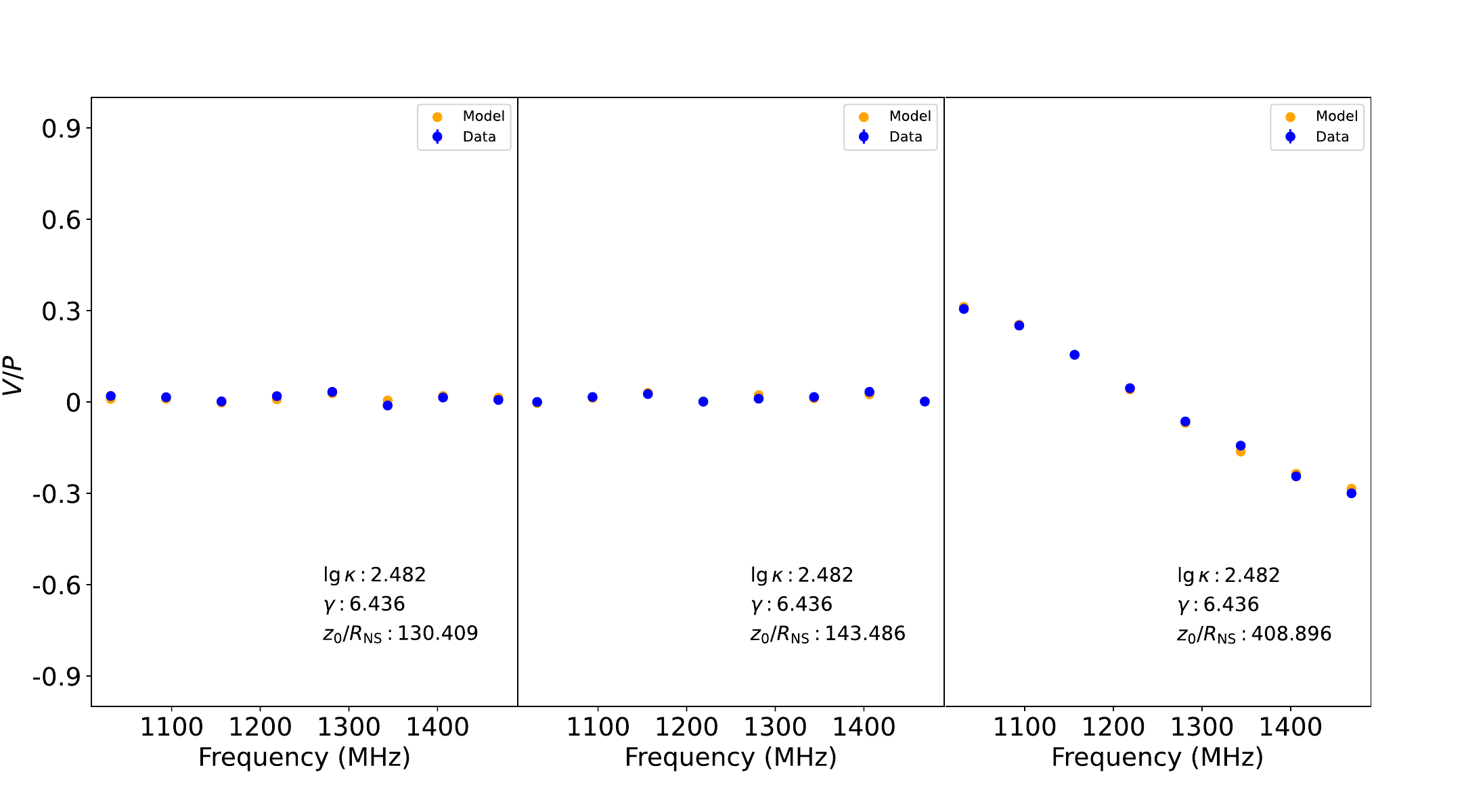}}
\end{minipage}

\vspace{0cm}
\begin{minipage}{\textwidth}
\centering
\includegraphics[width=8cm]{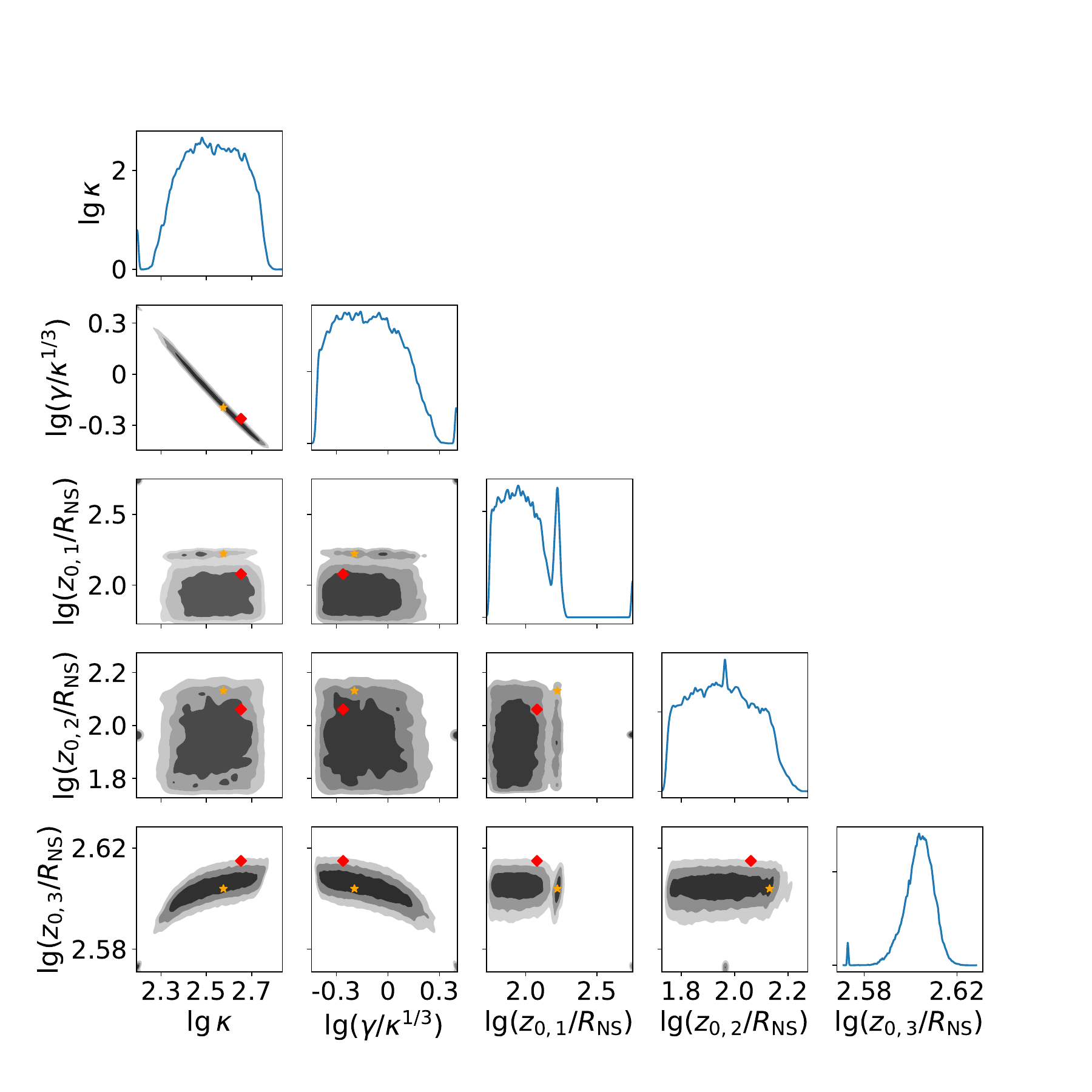}
\raisebox{0cm}{\includegraphics[width=8cm]{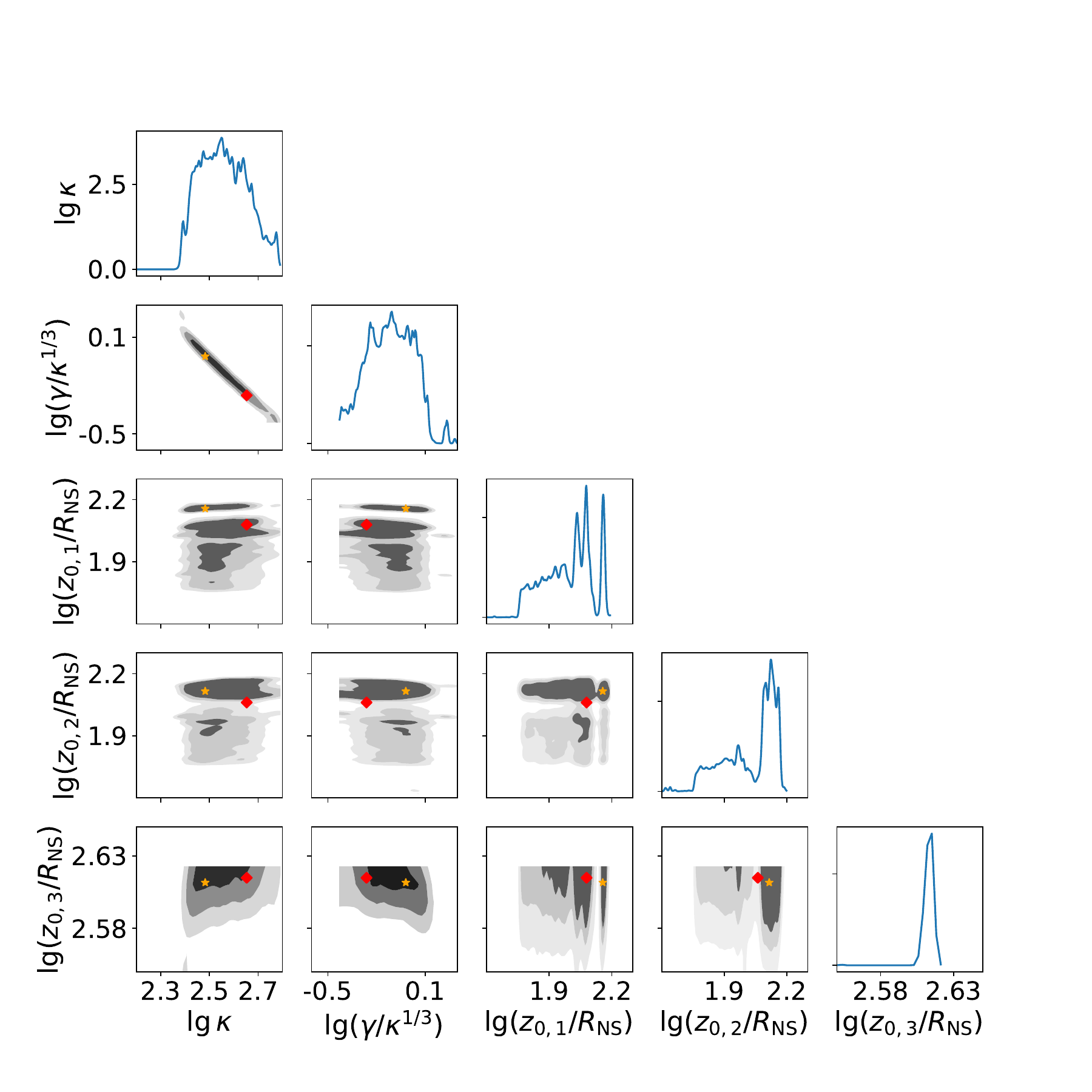}}
\end{minipage}
      \caption{Bayesian analysis results of a simulated sample with $\sigma=0.03$ (left) and $\sigma=0.01$ (right). The upper plots: simulated $V/P$s (blue dots with errorbars) v.s. modeled $V/P$s (orange dots) with maximum-likelihood parameters. The maximum-likelihood parameters are marked in the plots. The lower plot: posterior PDFs of five parameters. The meanings of plots are same as those in Figure~\ref{fig:contour_real}. The red diamonds represent the parameters set for simulation, and the orange stars represent the maximum-likelihood parameters.}
         \label{fig:simulated_003}
\end{figure*}

The $\chi_{r}^2$s of the fitting of simulated data lie within the range of (0.4, 1.2). Examples of the fitted curves v.s. simulated curves, and of posterior PDFs are shown in Figure~\ref{fig:simulated_003}. The fitting is generally good, with the fitted curves close to simulated curves. The posterior PDFs are sometimes multi-peaked, which reflects the degeneracy of parameters of the model.

\begin{figure*}[h!]
   \centering
    \includegraphics[width=12cm]{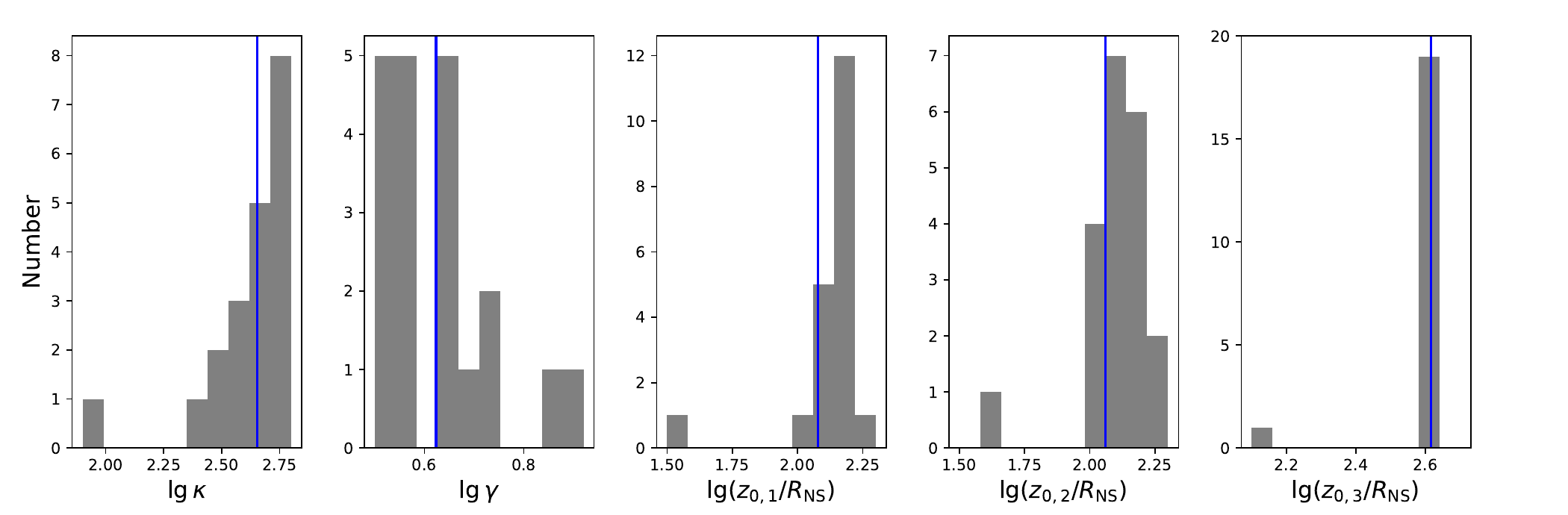}
    \includegraphics[width=12cm]{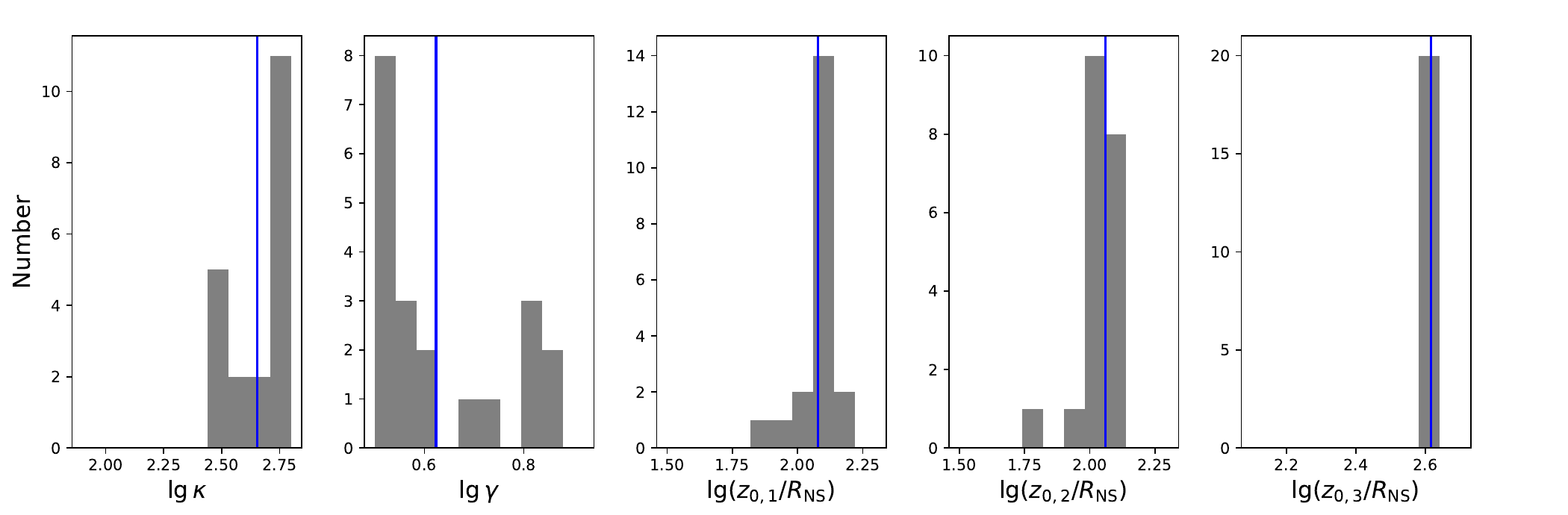}
      \caption{The distributions of maximum-likelihood parameters of 20 simulations, for $\sigma=0.03$ (upper) and $\sigma=0.01$ (lower). The blue vertical lines mark the reference parameters set for simulations.}
         \label{fig:simulation_paras}
\end{figure*}

The distributions of the maximum-likelihood parameters of 20 simulations are shown in Figure~\ref{fig:simulation_paras}. The mode coupling model seems to constrain the parameters differently. In the case of our simulation, the $\kappa$ and $\gamma$ are not as well constrained as $z_{0,(1,2,3)}$.

\section{Comparison of parameters of multi-peaks in posterior PDFs, and more examples of fitted pulses}\label{sec:comparison_app}
\begin{figure*}[h!]
   \centering
    \includegraphics[width=17cm]{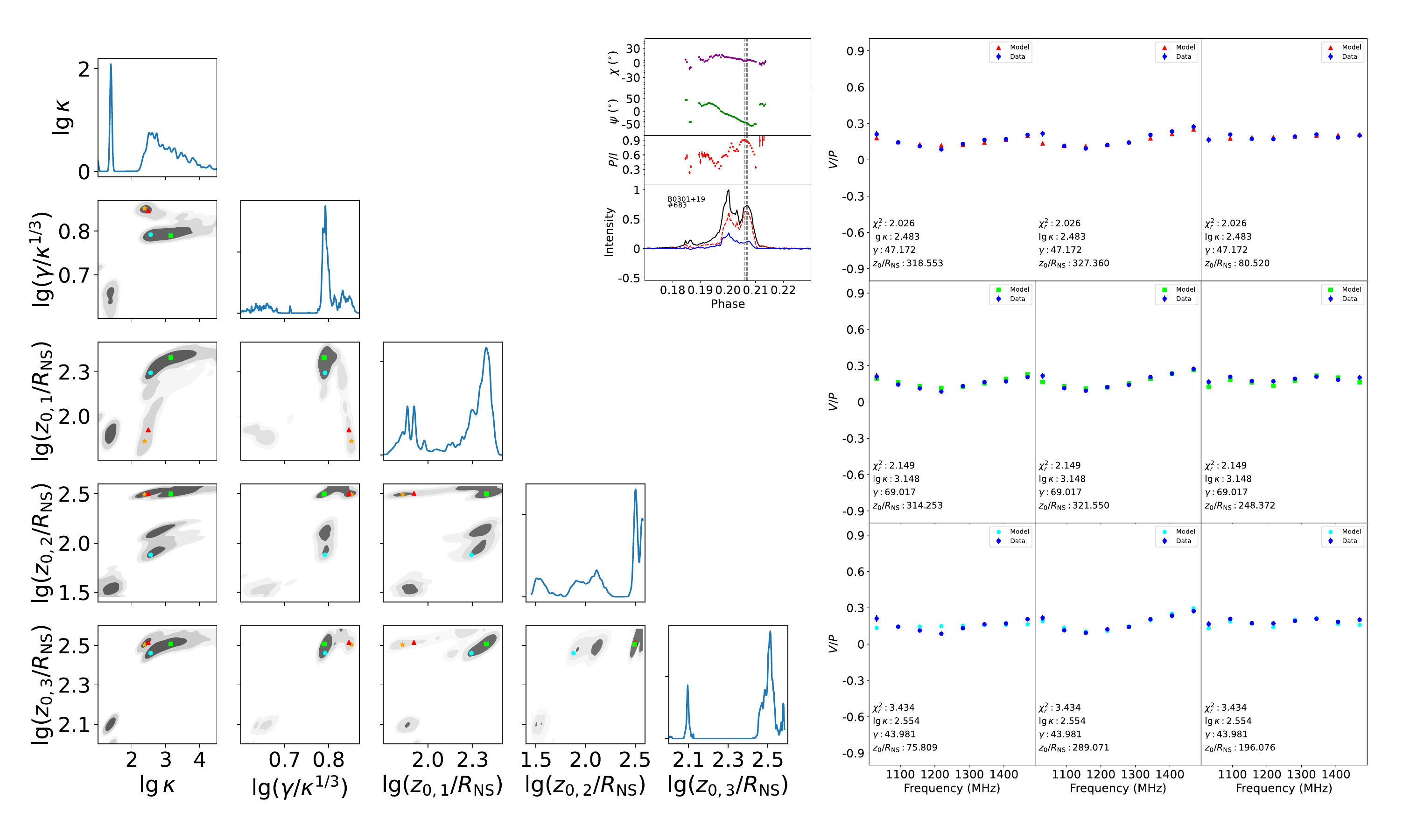}
    \caption{Left: Posterior PDFs of \#683 of B0301$+$19. The meanings of plots are same as those in Figure~\ref{fig:contour_real}, except that the red triangle, the green squares, and the cyan regular hexagon are markers of three 5D peaks chosen by using clustering package \texttt{hdbscan}. The comparison between real data (with errorbars) and modeled $V/P$ - $\nu$ curves (with corresponding parameters marked with the red triangle, the green squares, and the cyan regular hexagon). The pulse profile is also plotted here, and the meanings of lines and dots are same as those in Figure~\ref{fig:well_fitted1}.}
    \label{fig:multi_peaks}
\end{figure*}

\begin{figure*}[h!]
   \centering
    \begin{minipage}{\textwidth}
\centering
\includegraphics[width=8.9cm]{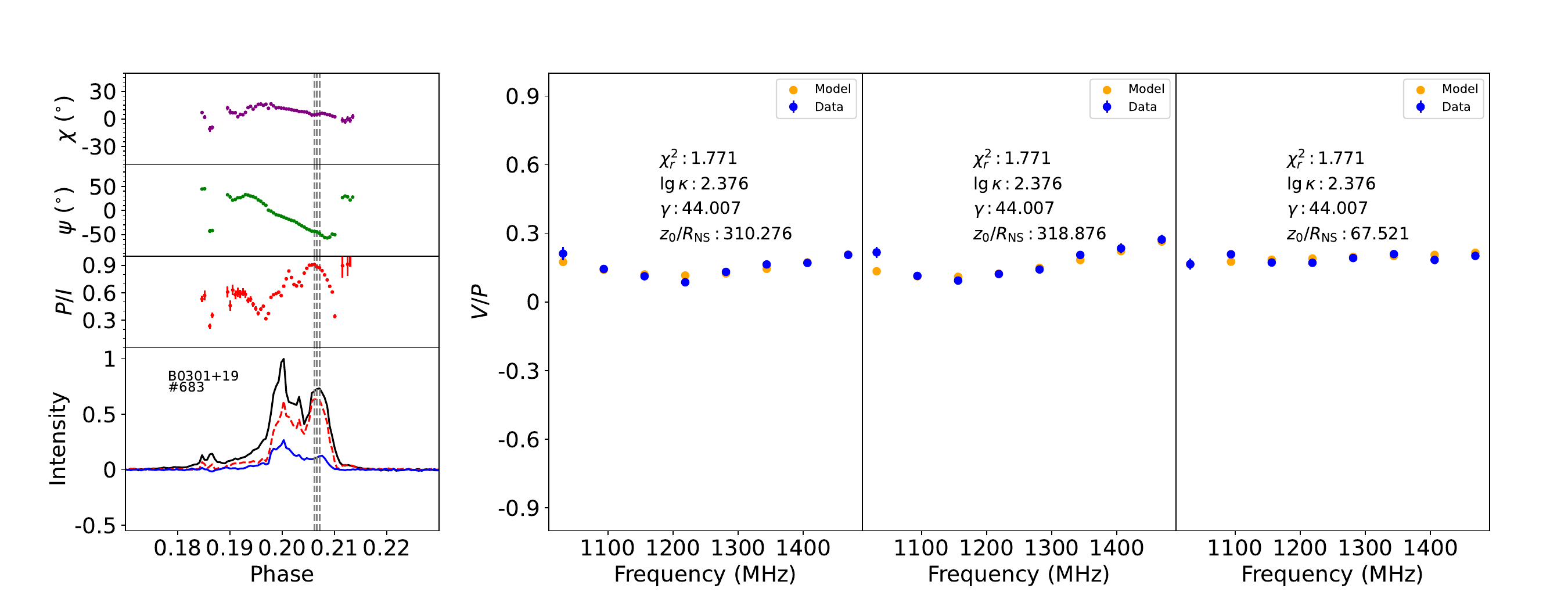}
\raisebox{0cm}{\includegraphics[width=8.9cm]{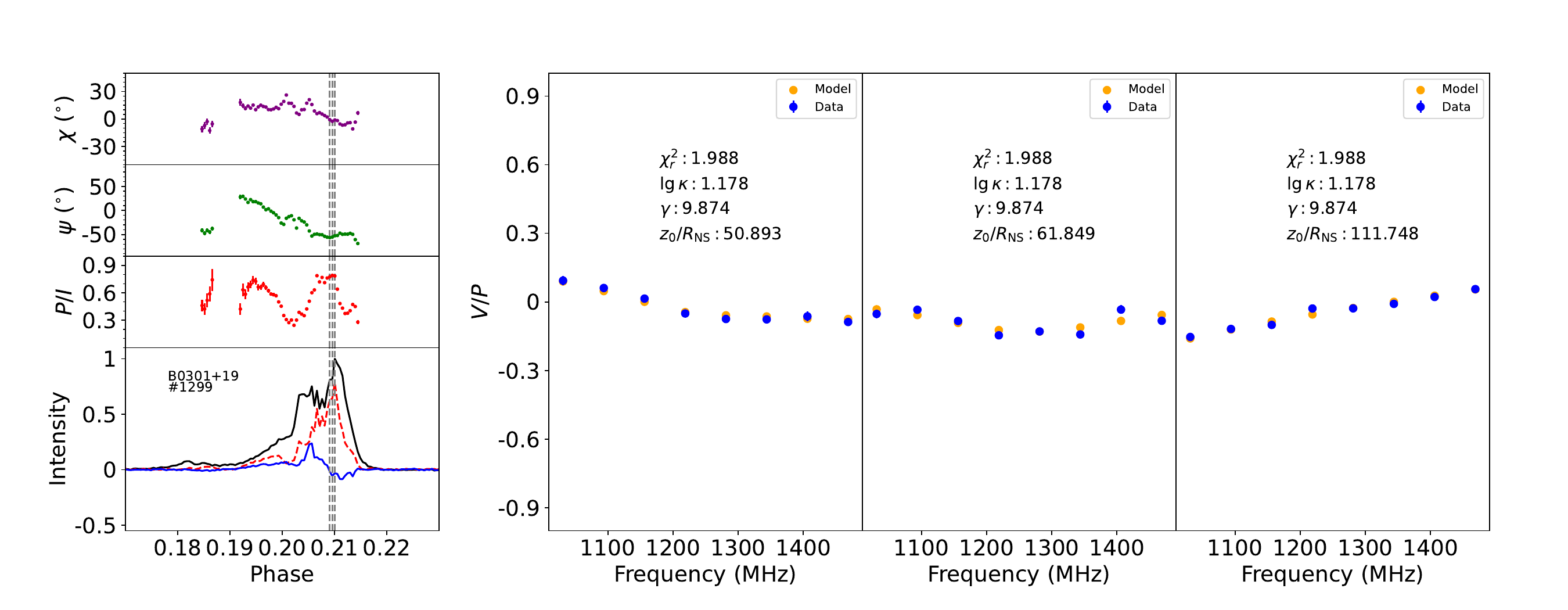}}
\end{minipage}

\vspace{0cm}
\begin{minipage}{\textwidth}
\centering
\includegraphics[width=8.9cm]{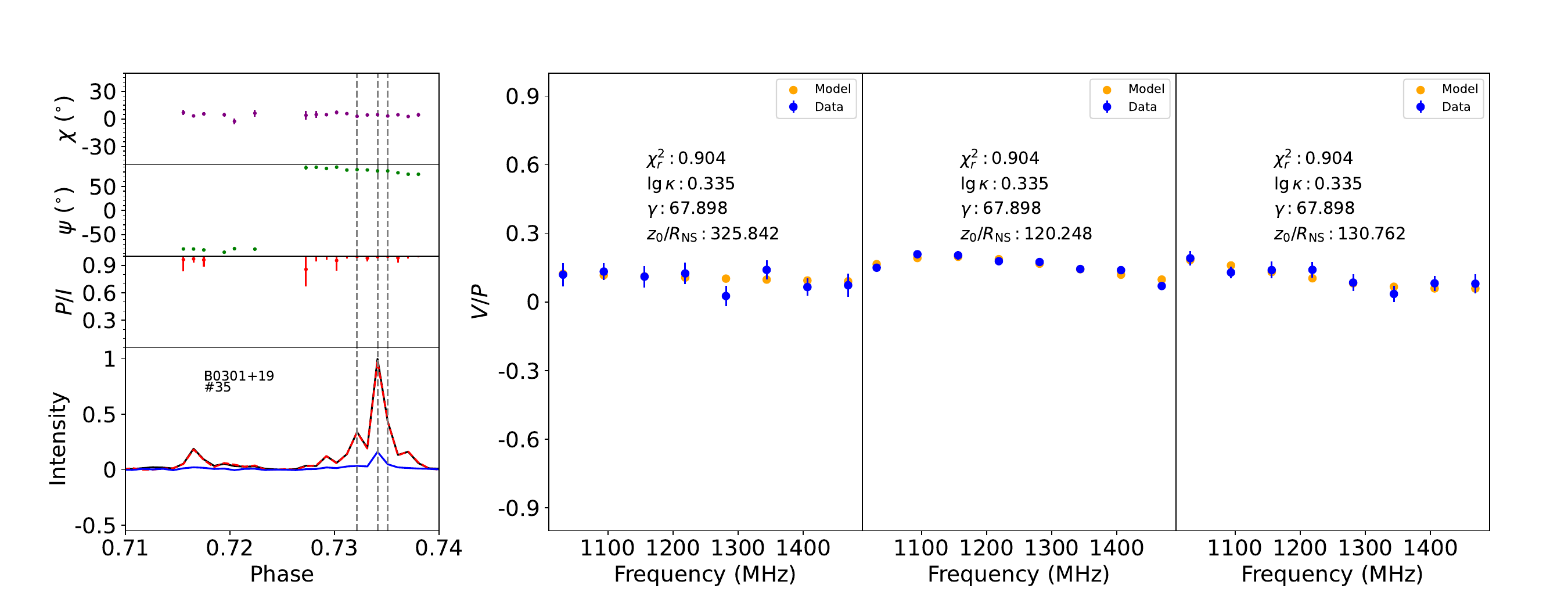}
\raisebox{0cm}{\includegraphics[width=8.9cm]{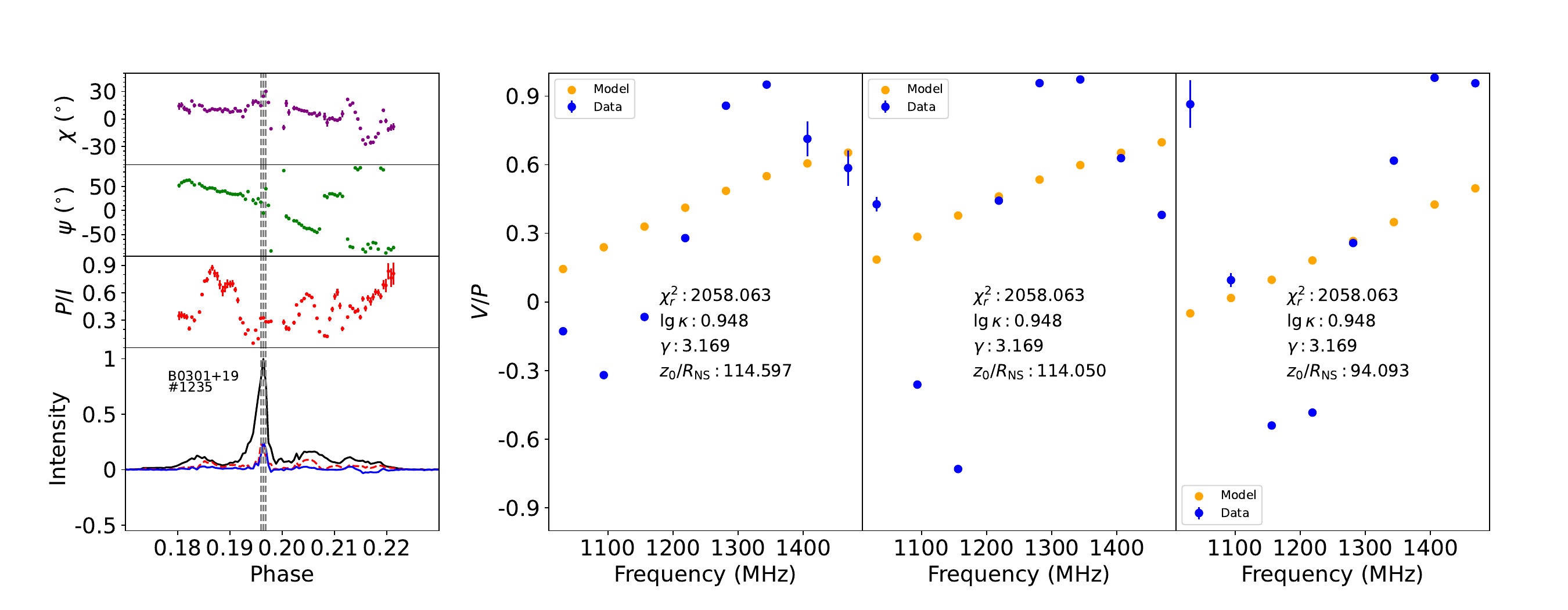}}
\end{minipage}

      \caption{Four pulses (12 $V/P$ - $\nu$ curves) with fitting results. The meanings of plots are same as those in Figure~\ref{fig:well_fitted1}.}
    \label{fig:more_examples}
\end{figure*}

The analysis of three peaks in the posterior PDF of \#683 of B0301$+$19 is shown in Figure~\ref{fig:multi_peaks}. Four pulses with fitted $V/P$ curves are shown in Figure~\ref{fig:more_examples}.

\end{appendix}
\end{document}